%% file: manuscript_arxiv.tex
\documentclass[iop]{emulateapj}
\usepackage{graphicx, wrapfig}
\usepackage{amsmath}
\usepackage{multirow}

\usepackage{natbib}
\newcommand{\captionfonts}{\normalsize}

\makeatletter  
\long\def\@makecaption#1#2{%
  \vskip\abovecaptionskip
  \sbox\@tempboxa{{\captionfonts #1 #2}}%
  \ifdim \wd\@tempboxa >\hsize
    {\captionfonts #1 #2\par}
  \else
    \hbox to\hsize{\hfil\box\@tempboxa\hfil}%
  \fi
  \vskip\belowcaptionskip
  }
\makeatother   

\newcommand{\galex}{\textit{GALEX}}
\newcommand{\msun}{\ensuremath{\text{M}_\odot}}
\newcommand{\ser}{S\'ersic}
\newcommand{\ugc}{UGC~1382}

\begin{document}

\hyphenation{LePHARE}

\title{On the Classification of UGC~1382 as a \\ Giant Low Surface Brightness Galaxy \\  \ }

\author{
Lea M.~Z.\ Hagen\altaffilmark{1,2}, 
Mark Seibert\altaffilmark{3}, 
Alex Hagen\altaffilmark{1,2}, 
Kristina Nyland\altaffilmark{4,5,6}, 
James D.\ Neill\altaffilmark{7}, 
Marie Treyer\altaffilmark{8}, 
Lisa M.\ Young\altaffilmark{4}, 
Jeffrey A.\ Rich\altaffilmark{3,9}
and 
Barry F.\ Madore\altaffilmark{3}
}

\affil{\altaffilmark{1}Department of Astronomy and Astrophysics, The 
Pennsylvania State University, University Park, PA 16802, USA} 
\affil{\altaffilmark{2}Institute for Gravitation and the Cosmos, The 
Pennsylvania State University, University Park, PA 16802, USA} 
\affil{\altaffilmark{3}Observatories of the Carnegie Institution for Science, 813 Santa Barbara Street, Pasadena, CA 91101, USA}
\affil{\altaffilmark{4}Physics Department, New Mexico Institute of Mining and Technology, Socorro, NM 87801, USA}
\affil{\altaffilmark{5}Netherlands Institute for Radio Astronomy (ASTRON), Postbus 2, NL-7990 AA Dwingeloo, the Netherlands}
\affil{\altaffilmark{6}National Radio Astronomy Observatory, Charlottesville, VA 22903, USA}
\affil{\altaffilmark{7}California Institute of Technology, Pasadena, CA 91125, USA}
\affil{\altaffilmark{8}Aix Marseille Universit\'{e}, CNRS, Laboratoire d'Astrophysique de Marseille, UMR 7326, 38 rue F. Joliot-Curie, F-13388 Marseille, France}
\affil{\altaffilmark{9}Infrared Analysis and Processing Center, California Institute of Technology, Pasadena, CA 91125, USA}


\begin{abstract}

We provide evidence that UGC~1382, long believed to be a passive elliptical galaxy, is actually a giant low surface brightness (GLSB) galaxy which rivals the archetypical GLSB Malin~1 in size.  
Like other GLSB galaxies, it has two components: a high surface brightness disk galaxy surrounded by an extended low surface brightness (LSB) disk.
For UGC~1382, the central component is a lenticular system with an effective radius of 6~kpc. Beyond this, the LSB disk has an effective radius of $\sim$38~kpc and an extrapolated central surface brightness of $\sim$26 mag/arcsec$^2$. Both components have a combined stellar mass of $\sim$$8 \times 10^{10}$ \msun, and are embedded in a massive ($10^{10}$~\msun) low density ($<3$~\msun/pc$^2$) HI disk with a radius of 110~kpc, making this one of the largest isolated disk galaxies known.
The system resides in a massive dark matter halo of at least $2 \times 10^{12}$~\msun.
Although possibly part of a small group, its low density environment likely plays a role in the formation and retention of the giant LSB and HI disks. 
We model the spectral energy distributions and find that the LSB disk is likely older than the lenticular component.
UGC~1382 has UV-optical colors typical of galaxies transitioning through the green valley.
Within the LSB disk are spiral arms forming stars at extremely low efficiencies. The gas depletion time scale of $\sim$$10^{11}$ year suggests that UGC~1382 may be a very long term resident of the green valley.
We find that the formation and evolution of the LSB disk in UGC~1382 is best explained by the accretion of gas-rich LSB dwarf galaxies.

\ \\

\end{abstract}



\section{Introduction} \label{sec-intro}

Giant low surface brightness (GLSB) galaxies are the most extreme low surface brightness (LSB) disk galaxies and are the largest isolated galaxies known to exist.  Although massive $(L \sim L^*)$ and gas rich ($\text{M}_\text{gas} > 10^{10}$~\msun), because they have disk scale lengths in excess of 10~kpc, they also have low gas surface densities and star formation efficiencies \citep{sprayberry95, impey97, matthews01}. Their rotation curves flatten near $V_\text{max} \sim$300~km/s and are dark matter (DM) dominated with DM fractions $> 0.7$ \citep{lelli10, buta11}. Despite the enormous size and luminosity of GLSB galaxies, their diffuse nature makes them difficult to detect and are assumed to be highly underrepresented in catalogs \citep{impey97}. Their contribution to the luminosity density of the universe remains unclear.  Their origins have implications for the success of $\Lambda$CDM and hierarchical formation at low densities.

GLSB galaxies are not simple `pure' low surface brightness systems. Rather, a defining characteristic is that they have both a normal high surface brightness (HSB) central component (typically an early type disk) which is embedded in a massive extended diffuse disk component \citep{lelli10, sprayberry95, barth07}.  Because star formation is usually present in the extended disks of GLSB galaxies \citep{boissier08}, they can be considered larger versions of the more recently defined category of Type~1 extended ultraviolet (XUV) disk galaxies \citep{thilker07}, in which UV emission is seen at distances well beyond the classical star formation threshold surface density.  GLSB galaxies, but for their large scale, are also similar to the  population of low-mass early type galaxies (i.e., elliptical and lenticular galaxies) that show low levels of star formation in the outer regions, which may be the result of recent accretion of lower mass galaxies
\citep{moffett12, salim10}.

There is no definitive formation scenario for GLSB galaxies, but most agree that a low density environment is required in order to build and keep such enormous, organized, tenuous, and seemingly undisturbed extended disks.  Although often described as simply unevolved gas-rich disks due to their low star formation efficiency \citep{bothun87, hoffman92}, the dual HSB inner region and LSB extended disk suggest a more complicated history which may involve both a rapid disk formation and a late collapse of a low-amplitude density perturbation \citep{impey97} or the tidal disruption of dwarf galaxies \citep[e.g.][]{penarrubia06}.  There are likely several mechanisms at work simultaneously. However they form, the relative isolation and low star formation efficiency suggests that they are not evolving rapidly at present.

The prototypical GLSB galaxy, Malin~1, discovered by \citet{bothun87}, has an extrapolated disk central surface brightness of $\mu_R(0) =$24.7 mag/arcsec$^2$, a staggering disk scale length of 57~kpc \citep[for $h = 70$;][]{moore06} and an absolute magnitude of $M_V = -22.9$ \citep{pickering97}. Malin~1's HI disk has a mass of $10^{11} M_\odot$ and extends to a radius of 110~kpc \citep{lelli10, pickering97}. Using Hubble Space Telescope imaging, \citet{barth07} confirmed that the inner 10~kpc of Malin~1 hosts a SB0/a disk of normal size and surface brightness. \citet{boissier08} has classified Malin~1 as having a Type~1 XUV disk.

Although more than a dozen systems are now considered to be GLSB galaxies \citep{matthews01, bothun90, sprayberry95}, no other system has been reported with properties as extreme as the prototypical Malin~1. In this article, we describe UGC~1382, which is nearly identical in terms of scale and other physical properties to Malin~1. However, at less than 1/4 the distance to Malin~1, it is significantly closer.  This allows a detailed multi-wavelength investigation of a true Malin~1-like GLSB galaxy at much smaller spatial scales, with the goal of constraining the formation and evolution of these extreme systems.

\ugc\ has been classified as an elliptical in many optical surveys
\citep{tonry81, laurikainen94, huchra99, doyle05, sanchez11, huchra12}.  
Several surveys looking for morphological features, such as stellar rings and bars, did not detect anything other than a simple bulge-dominated galaxy \citep{meyer04, nair10, baillard11}.
It has spectroscopically measured recession velocities ranging between 5550 and 5770~km/s 
\citep{huchra83, huchra99, meyer04, garcia09, aihara11}. 
We find a 21~cm systemic radial velocity of 5591~km/s (see \S\ref{sec-HI}) and adopt a distance of 80~Mpc \citep{wright06} in this paper; this gives a scale of about 380~pc/arcsec or 23~kpc/arcmin.
UGC~1382 may be in a small group; there are three known galaxies
within 1.5~Mpc.
UGC~1382 was found to have $5 \times 10^9$~\msun\ of HI gas \citep{garcia09}, 
which is approximately 13\% of the stellar mass
\citep{west10}.  
The only hint that it may be more noteworthy was the suggestion of an extended HI disk \citep{garcia09}, though no analysis of such a disk was undertaken.

UGC~1382 came to our attention during an investigation of star formation in early type galaxies.  We noticed that it contained a set of very extended spiral arms in ultraviolet (UV) imaging from the Galaxy Evolution Explorer \citep[\galex;][]{martin05}.
Further investigation revealed that this system is not an elliptical galaxy, but is in fact a GLSB galaxy composed of a HSB lenticular core and an 80~kpc radius LSB disk.
In order to better understand this unusual galaxy, we have assembled a set of multiwavelength data, ranging from radio to far-ultraviolet, which we present in Section~\ref{sec-data}.
In Section~\ref{sec-morph}, we discuss the galaxy's morphology, surface brightness profiles, HI gas content, star formation efficiency, LSB characteristics, and environment.
We derive the dark matter content of the galaxy in Section~\ref{sec-dm}.
We then use the multiwavelength photometric data to model the spectral energy distribution (SED) of the galaxy, its HSB lenticular component, and its extended LSB disk in Section~\ref{sec-model}.  
In Section~\ref{sec-green}, we examine the past and future evolution of UGC~1382 based on both its morphology and modeled physical parameters.
We present possible formation scenarios in Section~\ref{sec-formation}.
Finally, we summarize our results in Section~\ref{sec-summary}.  We use flat $\Lambda$CDM cosmology with $\Omega_\Lambda = 0.7$, $\Omega_\text{M} = 0.3$, and $H_0=70$~km/s/Mpc throughout.


\section{Data} \label{sec-data}

\begin{figure*}
	\centering
	\includegraphics[trim = 15mm 40mm 13mm 25mm, clip=true, width=0.9\textwidth, angle=180]{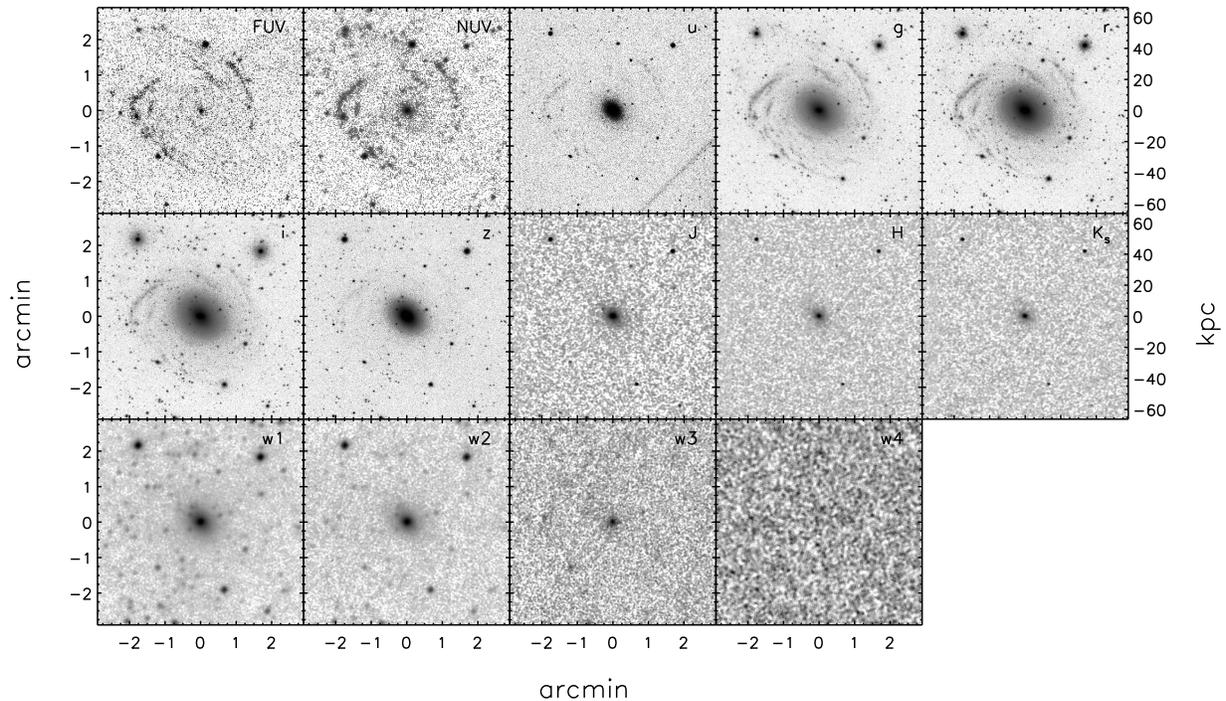}
	\caption{Multiwavelength images of \ugc. The morphology is dominated by the LSB spiral arms in the UV, whereas the central lenticular component is most dominant in the optical and IR.
	Each image is logarithmically scaled to bring out the most detail. The images are each $5.8'$ (130~kpc) wide.}

	\label{fig-panel}
\end{figure*}

\begin{figure*}
	\centering
	\includegraphics[trim = 5mm 5mm 10mm 10mm, clip=true, width=0.9\textwidth]{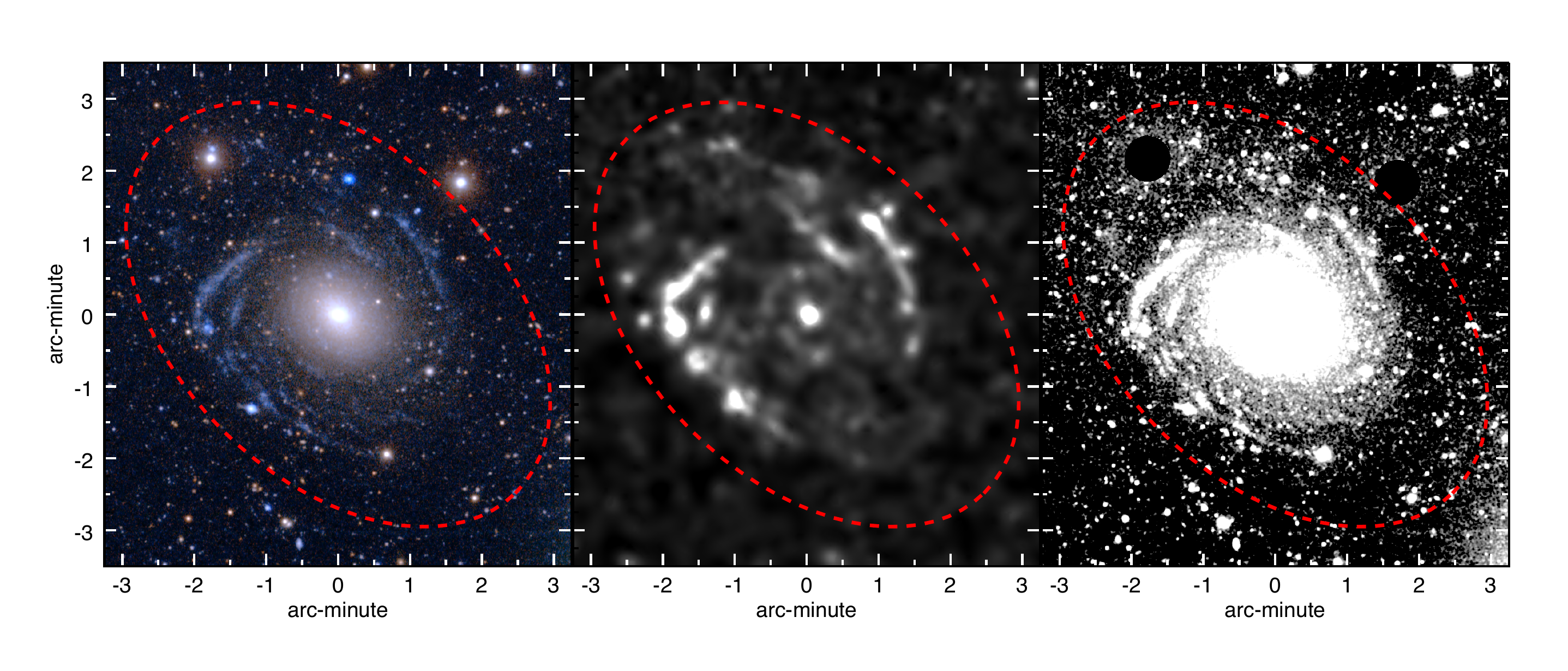}
	\begin{minipage}{0.9\textwidth}
	\caption[blah]{A multiwavelength view of the LSB disk and spiral arms in UGC~1382.  In each panel, the ellipse has a semi-major axis of 80~kpc and represents the largest radius at which we detect stellar light.
	\textit{Left}: Color composite image of UGC~1382.  The red channel is $i$, the green channel is $r$, and the blue channel is a combination of $g$ and NUV.  Patchy blue light to the northeast traces a faint spiral arm.
	\textit{Middle}: The FUV image adaptively smoothed\protect\footnotemark[1]\ to a signal-to-noise of 10 to reveal the extended LSB disk while minimizing foreground and background contamination.
	\textit{Right}: The $r$-band image smoothed with a 4~pixel ($1.6''$) boxcar kernel to highlight the patchy northeastern spiral arm.  The two bright stars to the northeast and northwest were masked prior to smoothing. 
	}
	\protect\footnotetext[1]{Adaptive smoothing utilized \texttt{asmooth}, described in ``Users Guide to the XMM-Newton Science Analysis System", Issue 11.0, 2014 (ESA: XMM-Newton SOC).}
	\end{minipage}
	\label{fig-color}
\end{figure*}

\ugc\ has been observed with 
\galex, the Sloan Digital Sky Survey (SDSS) Stripe 82 \citep[optical photometry and nuclear spectrum;][]{abazajian09, alam15},
the Two Micron All-Sky Survey \citep[2MASS;][]{cohen03}, 
and the Wide-field Infrared Survey Explorer \citep[WISE;][]{wright10}.  
Central wavelengths of each bandpass are listed in Table~\ref{tab-observations}.  Images of the galaxy in each of the fourteen filters between 0.15 and 22$\mu$m are shown in Figure~\ref{fig-panel}, and a color composite is shown in the left panel of Figure~\ref{fig-color}.

Photometry is performed and surface brightness profiles are generated using the WISE Nearby Galaxy Atlas (WNGA) and GALEX Large Galaxy Atlas (GLGA) pipeline (Seibert \& Neill, in prep).  Foreground stars and background galaxies were masked prior to analysis.  No $k$-corrections were made, as the galaxy is sufficiently local.  A summary of the photometry is in Table~\ref{tab-observations}.  The magnitudes shown in Table~\ref{tab-observations} are not corrected for the foreground galactic reddening of \mbox{\text{E(B$-$V) = 0.032}} \citep{schlegel98}, though corrections are applied prior to analysis.  The magnitudes are calculated within fixed apertures of sizes listed in Table~\ref{tab-observations}, which correspond to the HSB and LSB galaxy components.

In addition to the SDSS nuclear spectrum, an optical spectrum of an outer spiral arm was obtained with the 2.5-m du~Pont Telescope at Las Campanas Observatory using the Wide Field CCD Camera (WFCCD) in long-slit spectroscopy mode. The instrument was configured with a grism providing wavelength coverage of 3650\AA\ -- 8500\AA\ with 375~km/s FWHM resolution. The data were obtained on October 10, 2013, with an exposure time of 3$\times$1200 seconds at an airmass of 1.2.  The knot spectrum was extracted over $6.3''$ along the $1.65''$ slit, representing an area of 1.5~kpc$^2$ at the distance of UGC~1382.

\input{table1.tex}


\begin{figure*}
	\centering

	\includegraphics[trim = 0mm 0mm 0mm 0mm, clip=true, width=0.9\textwidth]{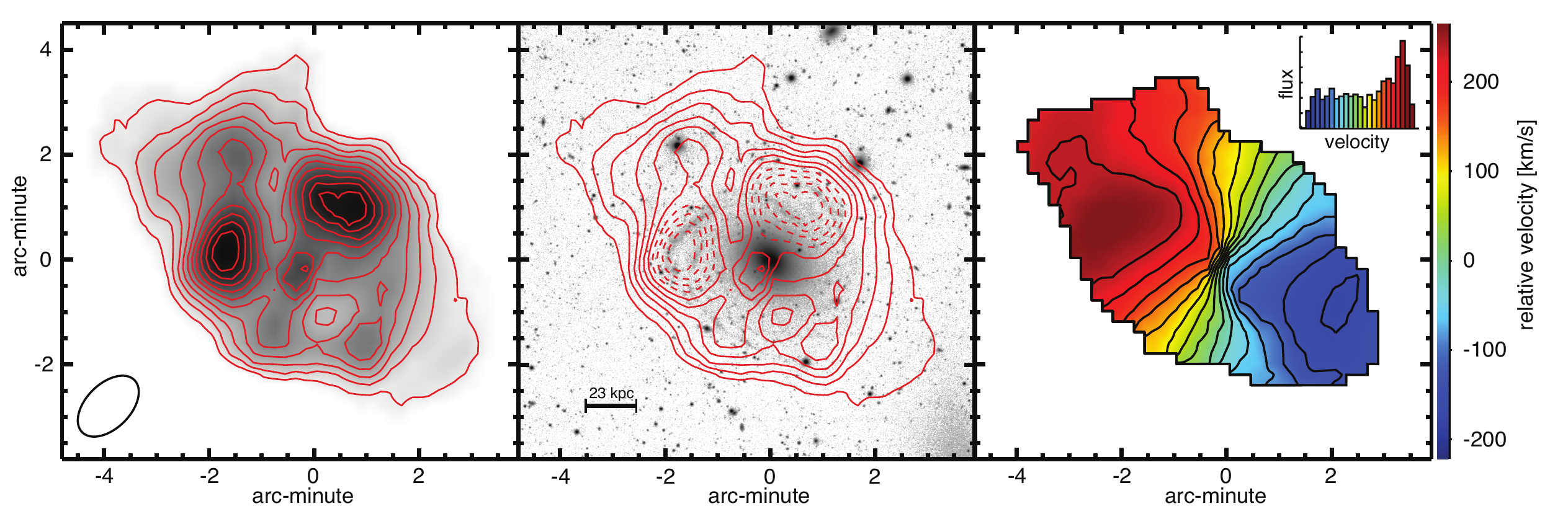}
	\caption{21~cm column density and velocity maps.  \textit{Left panel}: HI column density map.  The ten contours are spaced linearly from $2.9 \times 10^{19}$ cm$^{-2}$ to $2.9 \times 10^{20}$ cm$^{-2}$.  The synthesized beam is shown in the lower-left corner of the image.  \textit{Middle panel}: Optical $r$-band image (greyscale) with the HI column density contours overlaid in red.  The dashed lines represent the two HI knots described in \S\ref{sec-HI}.  \textit{Right panel}: Velocity map and spectrum. }
	\label{fig-radio}
\end{figure*}

Finally, UGC~1382 was previously observed with the NRAO\footnote{The National Radio Astronomy Observatory is a facility of the National Science Foundation operated under cooperative agreement by Associated Universities, Inc.} Very Large Array (VLA) in the D-configuration on March 31, 2007 (project ID: AY177).  The VLA correlator was configured with a total bandwidth of 6.3~MHz divided into 64 channels centered on HI at the systemic velocity of the galaxy.  The total integration time on the science target was three hours.  Observations of UGC~1382 were preceded and followed by observations of the phase reference calibrator, J0149+0555, every 30 minutes over a switching angle of 6$^{\circ}$.  The positional accuracy of the phase calibrator was $<0.002^{\prime \prime}$.  The calibrator 3C48 was used to set the amplitude scale to an accuracy of 3\% and calibrate the bandpass.  Data calibration and image processing were carried-out with the December 31, 2008 release of the Astronomical Image Processing System ({\tt AIPS}) following standard procedures.  Our final image cube has an rms noise of 0.7 mJy beam$^{-1}$ per channel, channel width of  97.66~kHz (20.52 km/s), and a synthesized beam with dimensions 76.93$^{\prime \prime} \times 50.86^{\prime \prime}$.  The HI column density map, velocity map, and spectrum are shown in Figure~\ref{fig-radio}.


\section{Physical Description and Classification} \label{sec-morph}

Despite its relatively small distance (80~Mpc) and relatively large angular size (3~arcmin), UGC~1382 is a clear case of a morphologically misclassified system. In the near-IR and optical, the HSB bulge component - so easily detected in shallow surveys - left the impression that UGC~1382 was a typical quiescent elliptical galaxy. As such, it has fallen into many samples classified as an elliptical.  The power of multiwavelength observations and deeper surveys reveals it to be a much more complicated system. 
It is unlikely that this is the only nearby system that suffers from such a misclassification.

In this section, we discuss the detailed UV/optical and HI morphology of UGC~1382.  We also address the broad characteristics of the optical spectra.  We calculate the star formation efficiency and place its LSB disk in the context of XUV disks.  Finally, we look at its galactic environment, and place the galaxy's physical size into context with other large galaxies.
Many of the quantities referenced here are summarized in Table~\ref{tab-summary}.

\input{table2.tex}

\subsection{Optical and Ultraviolet Morphology}

\begin{figure}
	\centering
	\includegraphics[trim = 5mm 40mm 50mm 0mm, clip=true, width=0.95\columnwidth]{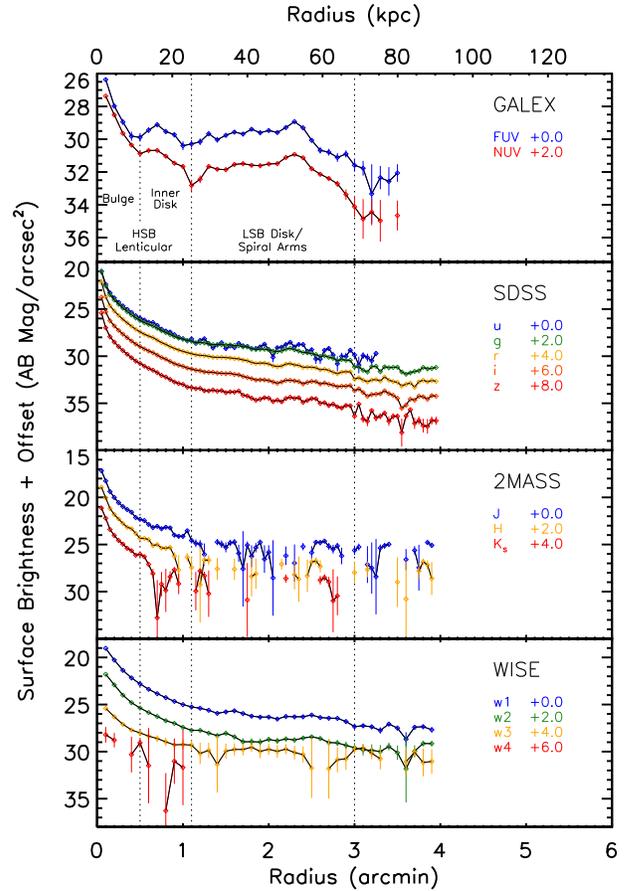}
	\caption{Radial surface brightness profiles of \ugc, divided (top to bottom) into UV (\galex), optical (SDSS), near-IR (2MASS), and mid-IR (WISE) panels.  The offset applied to each profile (in magnitudes) is listed at the right in each of the panels.  Dotted lines denote the divisions between the morphological components discussed in the text, and are identified in the top panel.}
	\vspace{2mm}
	\label{fig-profile}
\end{figure}

\begin{figure}
	\centering
	\vspace{5mm}
	\includegraphics[trim = 5mm 84mm 25mm 20mm, clip=true, width=0.99\columnwidth]{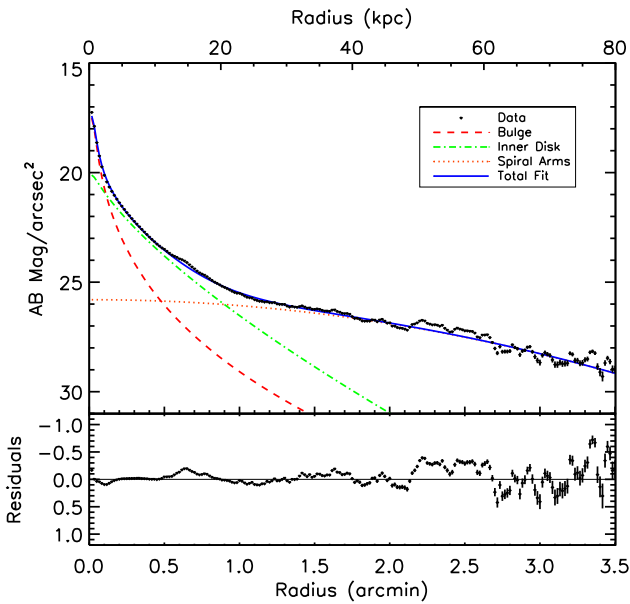}
	\caption{A three-component fit to the $r$-band surface brightness profile.  The bulge, inner disk, and outer LSB disk (which includes the spiral arms) have \ser\ indices of 3.5, 1.4, and 0.5, respectively.
	The bulge and inner disk together make the HSB component.}
	\vspace{2mm}
	\label{fig-profile_fit}
\end{figure}

A careful inspection of the annularly-averaged radial profiles (Figure~\ref{fig-profile}), along with a detailed decomposition of the $r$-band profile (Figure~\ref{fig-profile_fit}), suggests that \ugc\ consists of three morphological components: (1)~a classical HSB bulge embedded in (2)~a HSB inner disk, which also contains a small and weak set of spiral arms, all of which are surrounded by (3)~a very extended LSB disk with spiral arms.


To quantify the $r$-band surface brightness profile (Figure~\ref{fig-profile_fit}), we fit \citet{sersic63} functions using GALFIT \citep{peng02, peng10}.  
Specifically, we use GALFIT to fit a 2D image generated from the 1D radial profile with the assumption of a constant axial ratio and position angle.
We attempt both two- and three-component fits, each with an axial ratio of 0.69 and a position angle of 45$^\circ$.  For each fit, we calculate the Akaike information criteria \citep[AIC;][]{akaike74}.  The AIC is used to determine which model should be utilized, and is a dimensionless value taking into account both the goodness of fit and the number of model parameters.  We find that the three-component fit is favored by many orders of magnitude, and the relative likelihood of the two-component fit is essentially zero.  The three S\'ersic functions correspond directly to the three morphological components.

The central component of UGC~1382, which has been classified by numerous authors as an elliptical galaxy, is in fact composed of a classical bulge and disk.  Photometry for this inner component is in Column~5 of Table~\ref{tab-observations}.  
The bulge has a \ser\ index of $n=3.5$ and an effective radius of $r_e = 3.4''$ (1.3~kpc), while the inner disk has a \ser\ index of $n=1.4$ and an effective radius of $r_e = 15.8''$ (6.0~kpc).  The \ser\ fits are summarized in Table~\ref{tab-sersic}.  Comparing the fluxes of the two fits, the bulge-to-disk ratio is 0.70.  All of these values indicate that the central component is consistent with a lenticular galaxy.

\input{table3.tex}

\begin{figure}
	\centering
	\vspace{5mm}
	\includegraphics[trim = 10mm 45mm 10mm 60mm, clip=true, width=0.95\columnwidth]{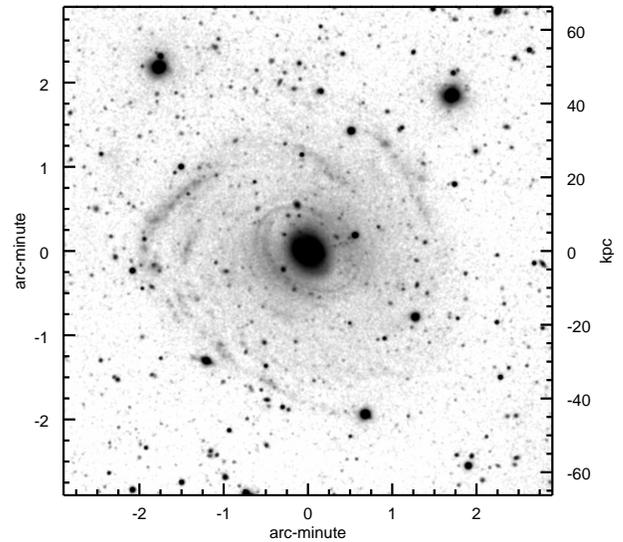}
	\caption{\ugc\ $r$-band image with the inner disk \ser\ component subtracted.  A set of tightly-wound spiral arms can easily be seen within a $1'$ radius of the center.}
	\vspace{2mm}
	\label{fig-disksub}
\end{figure}

In Figure~\ref{fig-disksub}, we subtract the \ser\ fit to the inner disk component, which clearly reveals a set of tightly wound spiral arms.  This feature is coincident with an inner ring-like structure seen in the far- and near-UV (FUV and NUV) images, which is also apparent in their radial profiles (labeled in the top panel of Figure~\ref{fig-profile}); therefore, these spiral arms are currently forming stars at a low level (discussed further in \S\ref{sec-model}).  We conclude that the HSB center of \ugc\ is a classic lenticular galaxy with evidence for weak spiral structure and recent star formation.  

Surrounding the normal lenticular galaxy, starting at a radius of $66''$ (25~kpc), there is an extended LSB disk with prominent spiral arms that we confidently detect to a radius of $3.5'$ (80~kpc).  Photometry for the spiral arms is in Column~6 of Table~\ref{tab-observations}.  
We note that the photometry extends to $3'$, but only about 2\% of the galaxy's light comes from the far outskirts, so conclusions we draw using the data in Table~\ref{tab-observations} are valid for the whole LSB component.
This component contributes the majority of the FUV and NUV flux (82\% and 70\%, respectively) from the galaxy.  Their blue color (FUV$-$NUV $=0.07$~mag) implies current star formation.  In the optical, these spiral arms are more difficult to detect with standard SDSS imaging, but with the deeper Stripe~82 imaging, we can convincingly quantify their flux. The full extent of the LSB disk and spiral arms can be seen in both the FUV and $r$-band imaging, shown in Figure~\ref{fig-color}.

 The \ser\ fit to the outer LSB disk in the $r$-band yields an index of $n=0.5$ and an effective radius of $r_e = 100''$ (38~kpc).  This is shallower than the exponential decline ($n=1$) typically seen in spiral galaxies.
Furthermore, the inner disk-subtracted image in Figure~\ref{fig-disksub} hints that the inner spiral arms may form a continuous structure with the outer spiral arms; we will return to this point in \S\ref{sec-formation}.

\begin{figure}
	\centering
	\includegraphics[trim = 5mm 55mm 35mm 60mm, clip=true, width=0.95\columnwidth]{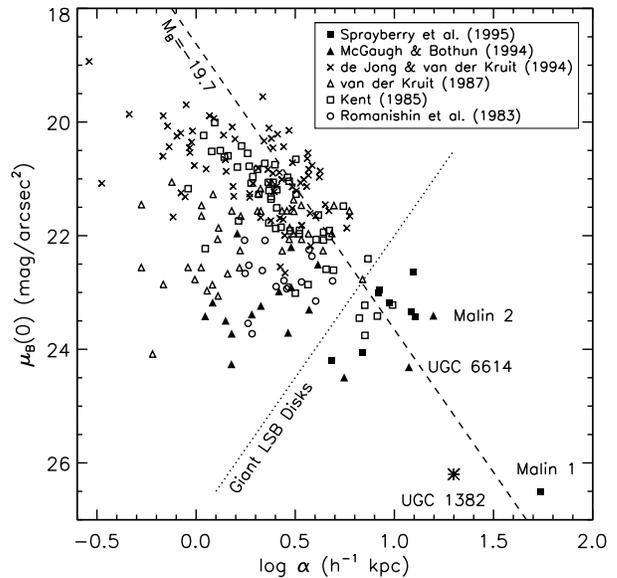}
	\caption{The relation between the B-band central surface brightness $\mu$ and scale length $\alpha$ of LSB disks, as recreated from \cite{sprayberry95}.  Galaxies with noteworthy GLSB disks are individually labeled. The dashed line represents a constant disk luminosity of $M_\text{B} = -19.7$ (corresponding to an $L^*$ disk), and the dotted line is the \citeauthor{sprayberry95} division between normal and GLSB disks.  UGC~1382 is clearly in the regime of GLSB disks.  
	\textit{Data sources}: \cite{romanishin83}, \cite{kent85}, \cite{vanderkruit87}, \cite{dejong94}, \cite{mcgaugh94}, \citet{sprayberry95}.}
	\label{fig-glsb}
\end{figure}

\begin{figure}
	\centering
	\includegraphics[trim = 20mm 105mm 35mm 30mm, clip=true, width=0.95\columnwidth]{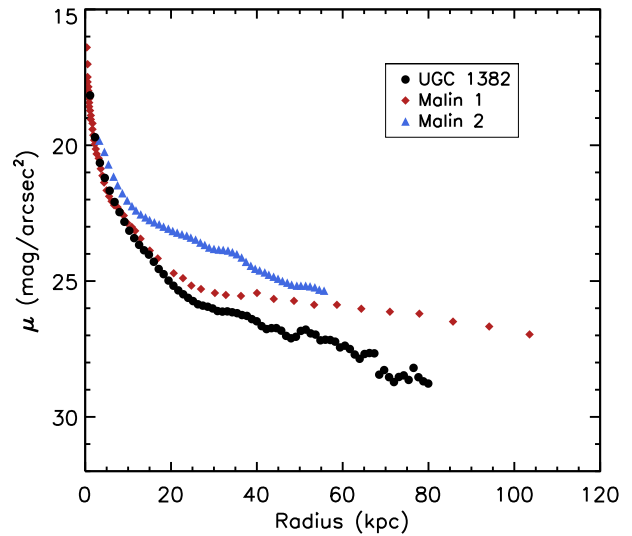}
	\caption{Comparison of the $r$-band radial profiles of UGC~1382, Malin~1, and Malin~2.  Both the physical extent and the extrapolated central surface brightness of UGC~1382 are similar to those of Malin~1.  UGC~1382 has an exponential scale length similar to that of Malin~2.
	\textit{Data sources}: \citet{barth07} (Malin~1, $r<10$~kpc), \citet{moore06} (Malin~1, $r>10$~kpc), \citet{kasparova14} (Malin~2)}
	\label{fig-malin}
\end{figure}

Finally, for comparison with the GLSB literature, we fit an exponential profile to the LSB disk for $r > 100''$.  We measure the extrapolated central surface brightnesses in the $g$- and $r$-band images and follow \citet{jester05} to transform the results to Johnson $B$-band.  This results in $\mu_B(0) = 26.2$ mag/arcsec$^2$ and an $r$-band scale length of $\alpha = 28.5 \pm 1.9$~kpc ($75''$).
\cite{sprayberry95} compare these same quantities for a variety of LSB disks and define a cutoff between normal and giant LSB disks using the ``diffuseness index," where GLSB systems have $\mu_B(0) + 5 \log \alpha > 27$ (for $h=100$).  The ``diffuseness index" of UGC~1382 is 32.7, so it is most certainly a GLSB galaxy.  This comparison, with the addition of UGC~1382, is shown in Figure~\ref{fig-glsb}, confirming not only that it is a GLSB galaxy, but that is the system most comparable to the extreme nature of Malin~1.
In addition, Figure~\ref{fig-malin} compares the radial profile of UGC~1382 to those of Malin~1 and Malin~2 \citep{bothun90}, the two other most extreme GLSB systems.  This also highlights that the exponential scale length of UGC~1382 is similar to that of Malin~2, whereas its physical extent and extrapolated central surface brightness are similar to those of Malin~1.


\subsection{Optical Spectroscopy}

\begin{figure*}
	\centering
	\includegraphics[trim = 0mm 2mm 0mm 0mm, clip=true, width=0.85\textwidth]{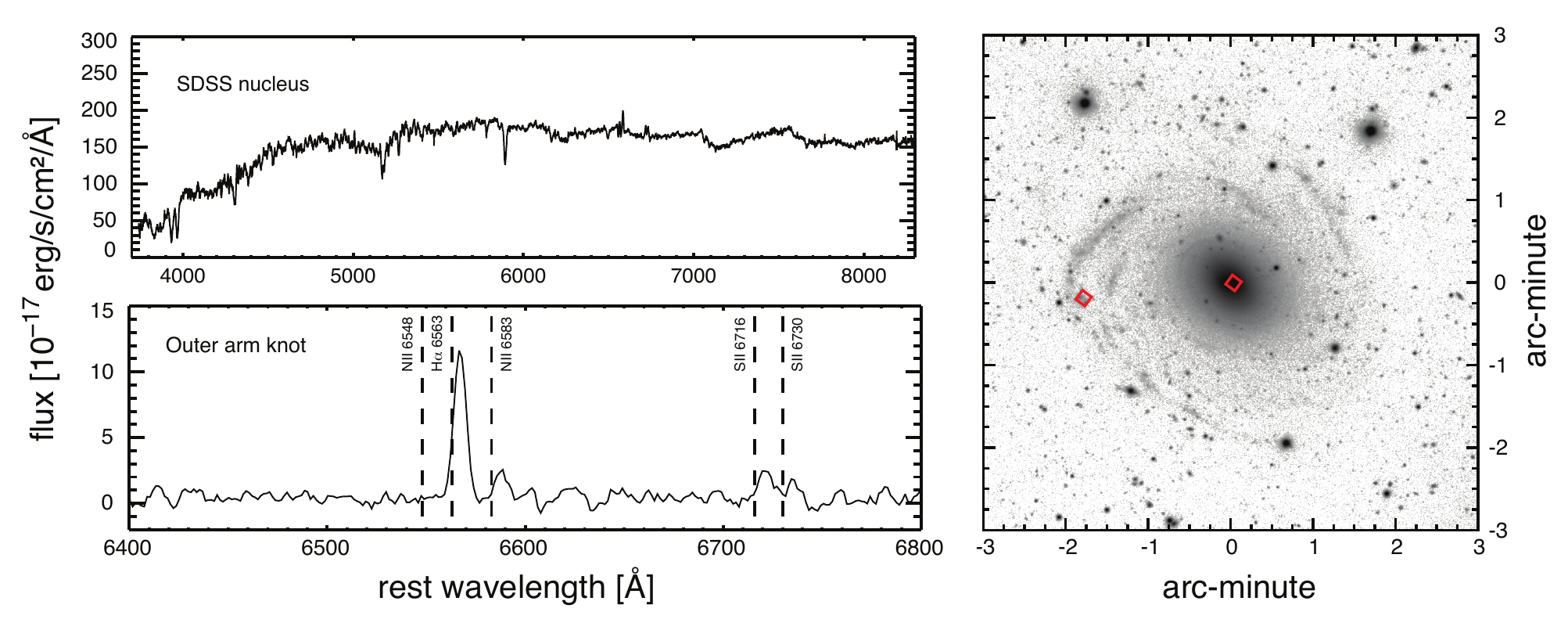}
	\caption{Spectra of the nucleus and a spiral arm/knot in UGC~1382.  \textit{Upper-left:} SDSS spectrum of the nucleus.  \textit{Lower-left:} Spectrum of a knot in the spiral arm.  Vertical lines denote the wavelengths of several lines at the galaxy's systemic velocity.  The +200~km/s velocity offset (with respect to the systemic velocity of UGC~1382) is consistent with the HI velocity at that location.  \textit{Right:} Image of UGC~1382 with the locations of the extracted spectra marked with red diamonds.}
	\vspace{2mm}
	\label{fig-spectra}
\end{figure*}

We have obtained spectra of one of the knots on the eastern side of the galaxy.  The resulting spectrum, shown in the bottom-left of Figure~\ref{fig-spectra}, displays a prominent H$\alpha$ line.  It is redshifted by about $+200$~km/s relative to the systemic velocity of the galaxy, and is consistent with the value of the HI velocity map at that position. 
The presence of the H$\alpha$ line confirms that the spiral arms are actively forming stars.

Also in Figure~\ref{fig-spectra} is the SDSS fiber spectrum ($r =1.5''$ or 570~pc) of the nucleus of UGC~1382.  It is typical of a bulge or early type galaxy.  The nucleus consists of an old stellar population, with no evidence for nuclear star formation or AGN activity.  The UV emission from the nucleus is, as expected, a result of old stellar populations \citep[e.g.,][]{brown00}.

\subsection{Neutral Hydrogen Content} \label{sec-HI}

The HI in UGC~1382, shown in Figure~\ref{fig-radio}, is distributed as a disk with a major axis of $9.6'$ (220~kpc) and an apparent axial ratio of 0.6.  It is centered on the UV/optical galaxy and is clumped in the two most prominent spiral arms.
The mass of the HI disk is calculated as
\begin{equation}
M_\text{HI} \ (\msun) = 2.36 \times 10^5 \ D^2 \int S(v) \ dv
\end{equation}
where $D$ is distance in Mpc and $\int S(v) \ dv$ is the integral of the line flux density in Jy~km/s.  We find an HI mass of $1.7 (\pm 0.1) \times 10^{10}$~\msun, which is a factor of 3 larger than the previous estimate of $5.6 \times 10^9$~\msun\ \citep{garcia09}.  We attribute this difference to our higher sensitivity and spatial resolution.  Due to the lower sensitivity of the \citet{garcia09} spectrum, the line fit to that spectrum only picks out the narrow peak that is visible at redshifted velocities in Figure~\ref{fig-radio}.

The HI column density map, seen in the left panel of Figure~\ref{fig-radio}, contains two bright knots, located to the east and northwest of the nucleus.  The knots together comprise 25\% of the HI flux.  The masses of the eastern and northwestern knots are approximately $1.8 \times 10^9$~\msun\ and $2.6 \times 10^9$~\msun, respectively.  As seen in the middle panel of Figure~\ref{fig-radio}, these knots are coincident with the brightest regions of the spiral arms, where the UV emission - and thus the star formation rate (SFR) - is the highest. 
The HI disk extends further to the northeast of the galaxy center than to the southwest, similar to the stellar disk highlighted in Figure~\ref{fig-color}.

The HI velocity map, shown in the right panel of Figure~\ref{fig-radio}, is consistent with a smoothly rotating disk with possible evidence for a slight warp.  The projected peak-to-peak velocity difference is $\sim$450~km/s.  The HI is clumped around the spiral arms, and there is a hint that the spiral arms may continue into the lenticular part of the galaxy (Figure~\ref{fig-disksub}), thus it is possible that the lenticular component may not be kinematically distinct from the spiral arms.  However, a more detailed analysis of the kinematics of the lenticular part would be necessary to confirm this.  The HI velocity map also rules out the presence of tidal streams, since we probe $-$500~km/s (blueward) and $+$800~km/s (redward) of the systemic velocity.

The velocity spectrum, displayed in the inset of Figure~\ref{fig-radio}, is slightly asymmetric.  
Ordinarily, this asymmetry indicates that the disk may be warped.  The eastern knot, which is located on the edge of the rotating disk, may also be enhancing the flux at the redshifted end of the spectrum.

Although the HI is concentrated in the vicinity of the spiral arms, the otherwise smooth distribution of HI gas and the uniformity of the HI velocity map imply that the HI disk is smoothly rotating and relatively undisturbed.  From this we infer that UGC~1382 has not recently been affected by a significant merger event.

\subsection{Star Formation Efficiency and XUV Disk Classification}

\begin{figure}
	\centering
	\includegraphics[trim = 25mm 130mm 25mm 30mm, clip=true, width=0.95\columnwidth]{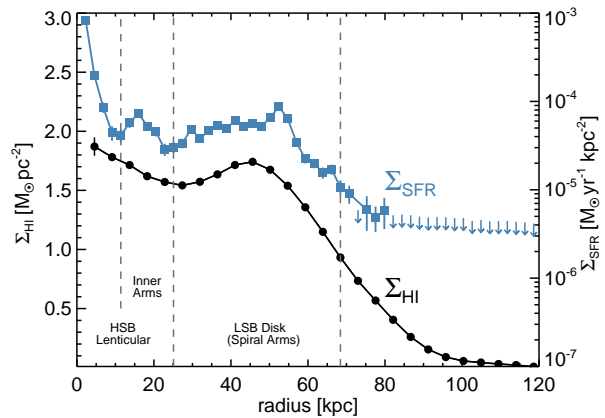}
	\caption{HI and SFR surface density profiles for UGC~1382.  Upper limits ($3\sigma$) are shown for data detected at less than $2.5\sigma$.  Morphological divisions are marked as in Figure~\ref{fig-profile}.  The two profiles follow each other closely.}
	\vspace{2mm}
	\label{fig-mass_sfr_profile}
\end{figure}

\newcommand{\sigmahi}{\ensuremath{\Sigma_\text{HI}}}
\newcommand{\sigmasfr}{\ensuremath{\Sigma_\text{SFR}}}

We compare the annular radial profiles of HI gas surface density (\sigmahi; not corrected to include helium) and the SFR surface  density (\sigmasfr) in Figure~\ref{fig-mass_sfr_profile}.   The annular averages of \sigmahi\ are less than $2~\msun/\text{pc}^2$, while the two HI knots (see Figure~\ref{fig-radio}) each have peak values of $\sim$3~$\msun/\text{pc}^2$.  The \sigmahi\ profile monotonically declines from the center to the edge of the inner disk (at $r=25$~kpc), then modestly increases again in the regions where the outer spiral arms become strong ($25 < r < 45$~kpc), and finally declines more rapidly beyond 45~kpc.

The \sigmasfr\ profile was derived from the FUV surface brightness profile, where the FUV has been corrected for intrinsic attenuation using the HI column density map \citep{bigiel10}. This attenuation correction takes advantage of the spatial distribution of the gas (and presumably dust) instead of assuming a single or annularly  averaged attenuation value. This yields a maximum attenuation of $A_\text{FUV} = 0.26$~mag within the HI knots and a mean (median) value of $A_\text{FUV} = 0.05$ (0.07)~mag over the entire HI disk.  We then performed annular photometry on the attenuation-corrected UV maps and applied a MW Galactic attenuation correction.  In order to derive the \sigmasfr\ profile shown in Figure~\ref{fig-mass_sfr_profile}, we used the \citet{salim07} FUV-to-SFR conversion.  This profile is very similar to the simple FUV profile shown in Figure~\ref{fig-profile} since the attenuation is so low.

The \sigmasfr\ profile mirrors the shape  of the \sigmahi\ profile fairly closely. The highest \sigmasfr\ also occurs at the galaxy's center and declines very rapidly over 10~kpc; however, this is likely due to a highly evolved population of low mass stars rather than recent star formation.  Within the disk of the lenticular component, the inner spiral arms cause a rise in \sigmasfr\ between $10 < r < 25$~kpc. A corresponding bump in \sigmahi\ is not seen in this region.  Beyond $r=25$~kpc, \sigmasfr\ increases modestly out to 50~kpc, about 5~kpc beyond the \sigmahi\ peak. This is also seen in the middle panel of Figure~\ref{fig-radio}, in which the spiral arms lead the HI peaks in the direction of rotation. We are able to reliably detect star formation out to 80~kpc, where $\sigmahi = 0.6~\msun/\text{pc}^2$.

Just like Malin~1, UGC~1382 also has much in common with the class of objects known as extended ultraviolet  (XUV) disk galaxies.
UGC~1382 can be classified as having a Type~I XUV disk, because it has structured UV-bright emission beyond the expected location of the star formation threshold ($\mu_\text{FUV} = 27.25$ mag/arcsec$^2$ or $\sigmasfr = 3 \times 10^{-4}$ \msun/yr/kpc$^2$), as defined by \citet{thilker07}.
The HI and SFR surface densities of UGC 1382 are typical of the low star formation efficiency found in the outer regions of spirals and dwarf galaxies \citep{bigiel10}.  In Figure~\ref{fig-gas_sfr}, we plot the distributions of pixel-by-pixel \sigmasfr\ and $\Sigma_\text{gas}$ of inner and outer regions of spiral galaxies from \citet{bigiel10}.  The red contours are the distribution of values for $r < r_{25}$ and the blue contours represent the distribution for $r > r_{25}$. The radial annular values of UGC~1382 for $r > 10$~kpc (the regions where FUV represents star formation) lie completely within the outer region distribution of{\citet{bigiel10}.  Even the lenticular disk of UGC~1382 is consistent with the outer regions of spirals; this is not surprising given the generally low SFR and gas density throughout UGC~1382.
The upturn of our data points at $\log \Sigma_\text{gas} = 0.35$ is due to the fact that our HI data have lower spatial resolution than do the spirals in \citet{bigiel10}.

\begin{figure}
	\centering
	\includegraphics[trim = 70mm 105mm 45mm 55mm, clip=true, width=0.95\columnwidth]{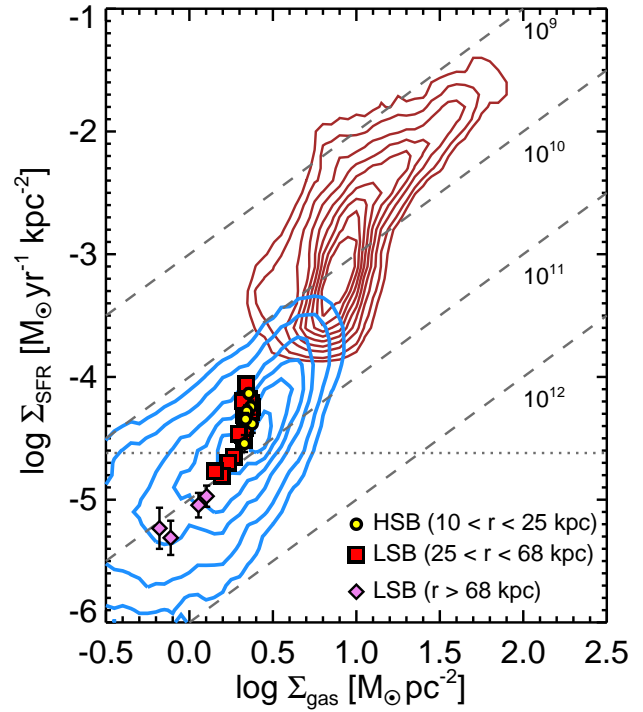}
	\caption{Comparison of the SFR and gas surface densities for UGC~1382.  Data points are from $6''$-wide annuli in the galaxy, divided into whether they came from the inner disk (yellow circles), outer LSB spiral arms (red squares) or beyond the spiral arms (purple diamonds).  The gas surface density is calculated by scaling the HI surface density by a factor of 1.36.  Contours represent the distributions for nearby spiral galaxies from \citet{bigiel10}: the upper (red) contours are for $r<r_{25}$ and the lower (blue) contours are for $r_{25}<r<2r_{25}$. The $3\sigma$ detection limit for the contour data is marked with a dotted line.  The dashed lines indicate gas depletion time scales of $10^8$ to $10^{12}$ years.}
	\label{fig-gas_sfr}
\end{figure}

\subsection{Environment}

Environment plays a significant role in the evolution of galaxies, and in this section, we review the details about the environment in which UGC~1382 resides.  To do this, we use the NED (NASA/IPAC Extragalactic Database) environment search tool\footnote{\url{https://ned.ipac.caltech.edu/forms/denv.html}} to find all  galaxies within 7.3~degrees (10~Mpc at the distance of UGC~1382) and $\pm$2000~km/s within the NED holdings.  One must keep in mind that the numerous surveys compiled by NED are not homogeneous, so it is likely that this set of galaxies is incomplete.

Figure~\ref{fig-env} shows the distribution of nearby galaxies in both physical and velocity space.  The first panel (top-left) represents the physical extent of its LSB disk with a radius of 100~kpc and a velocity difference of $\pm$250~km/s.  One neighboring galaxy satisfies these conditions.
The second panel increases the physical separation to 200~kpc and the velocity difference to $\pm$500~km/s, and yields an additional two neighboring systems.

\begin{figure*}
	\centering
	\includegraphics[trim = 15mm 12mm 15mm 15mm, clip=true, width=0.85\textwidth, page=1]{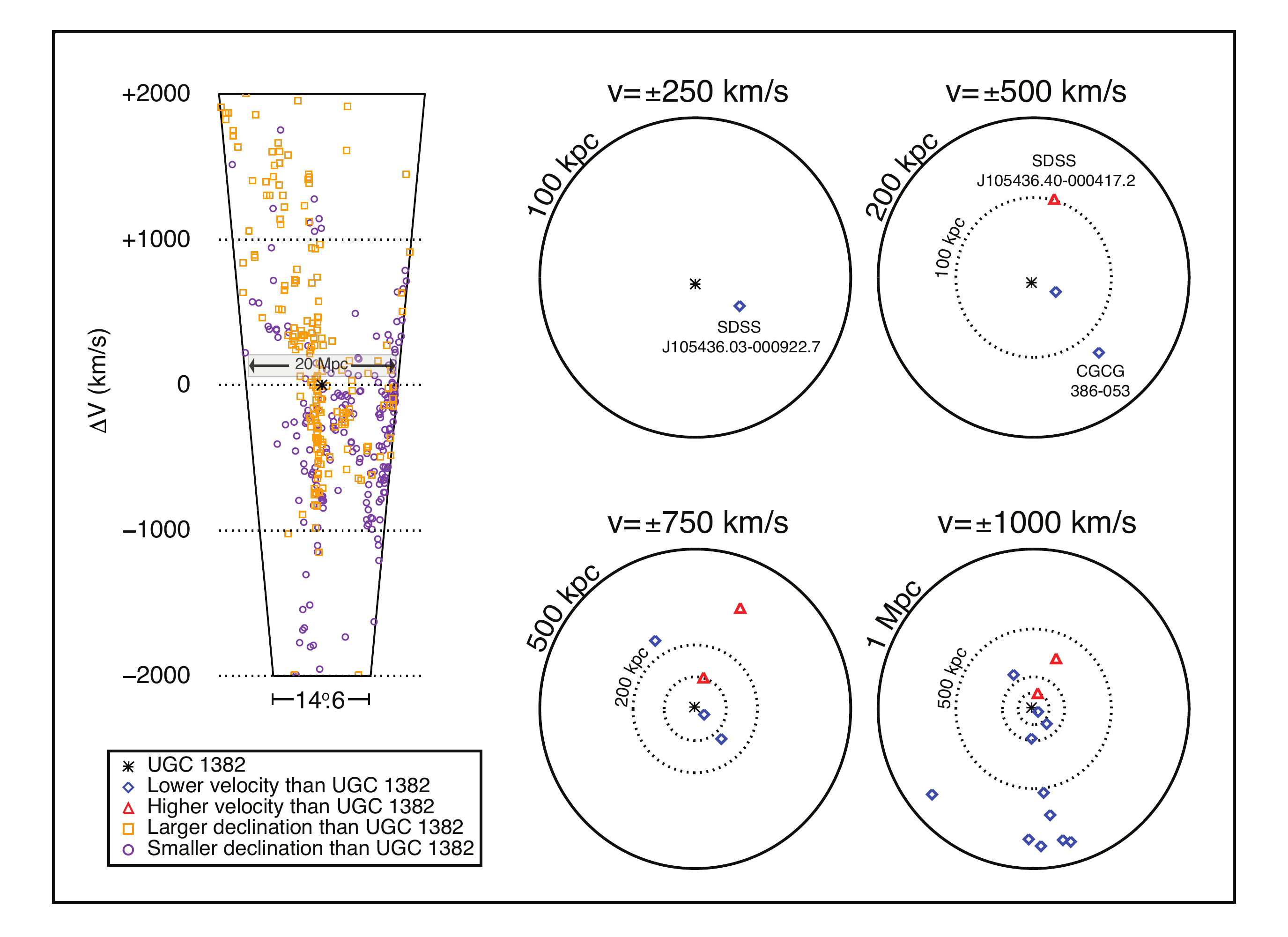}
	\caption{The environment of UGC~1382. 
		\textit{Cone}: Galaxies with radial velocities within $\pm$2000~km/s of UGC~1382 and within a 7.3$^\circ$ radius (10~Mpc at UGC~1382).  Galaxies with higher declination than UGC~1382 are marked with orange squares, and those with lower declination are marked with purple diamonds.  This view is much wider than those in the circles, and shows a larger-scale view of the galaxy's environment.
		\textit{Circles}: Galaxies within 100~kpc to 1~Mpc projected separation and within $\pm$100~km/s to $\pm$1000~km/s of the radial velocity of UGC~1382.  Dotted lines are the distances from the previous circle.  Galaxies with smaller radial velocities than UGC~1382 are marked with blue diamonds, and those that have larger radial velocities are marked with red triangles.  Galaxies are positioned based on their location in the plane of the sky.  The upper-left circle has radial and velocity cuts corresponding to the size and rotation of the LSB disk of UGC~1382; the single neighboring galaxy is likely embedded in the LSB disk.}
	\vspace{2mm}
	\label{fig-env}
\end{figure*}

These three closest galaxies warrant further discussion.  The closest, SDSS~J015436.03-000922.7, is a tiny red galaxy that is projected within the LSB disk of UGC~1382.  It was targeted by SDSS despite its faint r-band magnitude ($m_r = 18.4$), as it was originally classified as a high-redshift quasar.  Its spectrum is suggestive of an old stellar population typical of a bulge.  Its morphology is a simple spherical bulge, and following \citet{bell03}, we estimate a stellar mass of $5 \times 10^8$~\msun.
It has an $r$-band Petrosian radius \citep{blanton01} of $2.69''$ (1~kpc), which makes it either a dwarf galaxy or the tidally stripped core of a larger system. It is 1.5~arcmin (34~kpc) to the southwest of the center with a radial velocity of 5495~km/s, which is 100~km/s blueward of the systemic velocity of UGC~1382.  This relative velocity is consistent with the rotation velocity of the HI disk at that location.  We therefore believe that the galaxy (or remnant) is embedded within the LSB disk.  Its physical significance is discussed further in Section~\ref{sec-formation}.

Two galaxies, SDSS~J015436.40-000417.2 and CGCG~386-053, have projected distances of 100~kpc and 130~kpc with velocity differences of $+95$~km/s and $-95$~km/s, respectively.  The former is a small galaxy, with a Petrosian radius of $r = 3.4''$ (1.3~kpc) and stellar mass of $10^9$~\msun, and likely hasn't played a significant role in the evolution of UGC~1382.  The latter, however, is significantly larger, with a radius of $11.5''$ (4.4~kpc) and a stellar mass of $1.4 \times 10^{10}$~\msun. This is just under 20\% of the UGC~1382 stellar mass, so it is plausible that CGCG~386-053 has influenced the evolution of UGC~1382.

We quantitatively measure the environment of UGC~1382 following \citet{blanton09}.  They use three criteria to select neighboring galaxies: a velocity within $\pm$600~km/s, a projected distance within $500/h$~kpc, and a brightness $M_R - 5 \log h < -18.5$.  The number of galaxies $N$ that meet these criteria indicates whether the galaxy is isolated ($N=0$), in a poor group ($1 \le N \le 3$), in a rich group ($4 \le N \le 9$), or in a cluster ($N \ge 10$).  Around UGC~1382, only CGCG~386-053 has the required proximity and brightness, suggesting that it is in an $N=1$ poor group.
We conclude that UGC~1382 is in a low-density environment, possibly in a group with CGCG~386-053.  Furthermore, there is likely a small bulge-like system within its LSB disk.

\subsection{The Physical Extent of UGC~1382 in Context}

It is interesting to compare the size of UGC~1382 to that of other giant galaxies.  As summarized in Table~\ref{tab-summary}, UGC~1382 has HI gas disk measured out to a diameter of 220~kpc, with optical light detected to a diameter of 160~kpc.

The largest known galaxy is IC~1101, a cD galaxy in Abell 2029, for which \citet{uson90} measured an R-band diameter of $\sim$600~kpc.  However, as a cD galaxy, IC~1101 is surrounded by tidal debris from the accretion of a multitude of smaller galaxies; it is not clear how to separate this intra-cluster light from the extended envelope of the galaxy \citep{covone06}.  Therefore, it is likely that the diameter is much smaller than 600~kpc, but still considerably larger than UGC~1382.

NGC~262 is a tidally disturbed spiral galaxy with HI dimensions of 216~kpc $\times$ 274~kpc \citep[adjusted to $h=0.7$;][]{simkin87}.  Due to the tidal interactions, it is not in a state of equilibrium, and it is therefore difficult to compare its size to an undisturbed system.

The optical disk of Malin~1 extends to a diameter of 220~kpc, with HI also detected to 220~kpc.  There is evidence that the outer LSB component is the result of interaction with neighboring galaxies \citep{mapelli08, reshetnikov10}. 
Malin~1 is currently considered to be the largest disk galaxy.  Malin~2 is a GLSB galaxy of similar scale to Malin~1, with an optical diameter of 120~kpc and HI diameter of 220~kpc, and may also be interacting with a low-mass satellite \citep{kasparova14}.
Due to the similar size of UGC~1382 and NGC~262, Malin~1, and Malin~2, we conclude that UGC~1382 is the among the largest known disk galaxies.


\section{Dynamical Mass and Dark Matter} \label{sec-dm}

We have constructed a rotation curve from the HI velocity map using \verb=velfit= \citep[version 2.0;][]{spekkens07, sellwood10}.  The velocity map has $12''$ (4.6~kpc) pixels, so we cannot accurately probe the inner 10~kpc of the rotation curve.  Outside of this radius, we measure the rotation curve at 3-pixel (14~kpc) increments, which is approximately the same scale as the beam size.  We do not attempt to model the disk as a warped disk.  The resulting curve is plotted in Figure~\ref{fig-rotation_curve}.

In order to measure the dark matter profile, we must first account for the mass profiles of stars and gas.
We calculate the mass of the stellar component using the $r$-band surface brightness profile.  We assume a radius-independent mass-to-light ratio of 2.66, which we calculated using the modeled stellar mass in Table~\ref{tab-sed} (see \S\ref{sec-model}) and the total $r$-band light in Table~\ref{tab-observations}.  
For the mass contribution of the gas, we scaled the HI mass profile by 1.36 to account for helium; regardless of this factor, however, the gas only constitutes a tiny fraction (1\%) of the system's total mass.

We fit the remaining dark matter with NFW \citep{navarro97} and Einasto \citep{einasto65} profiles.  The best fits are shown as dashed blue and red lines, respectively, in Figure~\ref{fig-rotation_curve}.  The Einasto profile is a marginally better fit than the NFW profile, as it captures the decreasing velocity at large radius.  We find that UGC~1382 is already dark-matter dominated at a radius of 5-10~kpc, just outside the bulge component of the lenticular portion of the galaxy.
At the outermost point in our measured rotation curve (110~kpc), we find that the total enclosed mass is $2 \times 10^{12}$~\msun, which corresponds to a dark matter fraction of 0.95 and an $r$-band mass-to-light ratio of $\sim$65.

\begin{figure}
	\centering
	\includegraphics[trim = 30mm 130mm 20mm 35mm, clip=true, width=0.95\columnwidth]{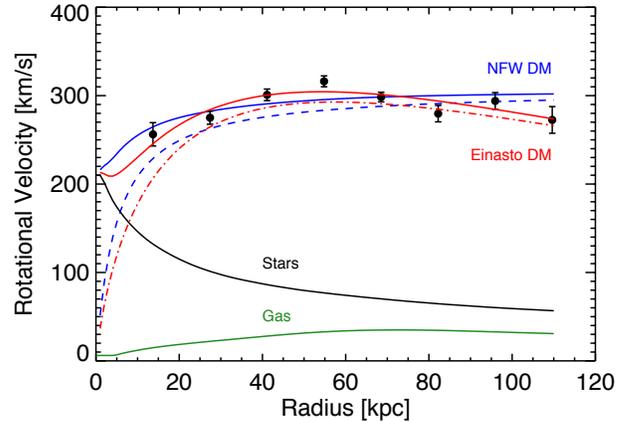}
	\caption{HI-derived rotation curve and contributing mass components.  The black points are the measured rotation curve.  The measured stellar and gaseous components are black and green lines, respectively.  Two models of the remaining dark matter - NFW and Einasto profiles - are shown in blue and red dashed lines.  The combined rotation curves, using each of the dark matter models, are plotted with solid blue and red lines.}
	\vspace{2mm}
	\label{fig-rotation_curve}
\end{figure}


\section{SED Modeling} \label{sec-model}

We now use spectral energy distribution (SED) fitting to explore the physical parameters of \ugc.  We fit the SEDs of the whole galaxy as well as those of the HSB lenticular and LSB spiral arm components.  To check the robustness of our physical parameter results, we have utilized two different fitting codes: GalMC \citep{acquaviva11} and LePHARE \citep{arnouts99, ilbert06}.  By using two fitting procedures, we can make better estimates of the physical parameters \citep{santini15, hayward15}, which allows us to infer the past and future evolution of the system.

\begin{figure*}
	\centering
	\includegraphics[trim = 25mm 25mm 30mm 20mm, clip=true, scale=0.7, angle=180]{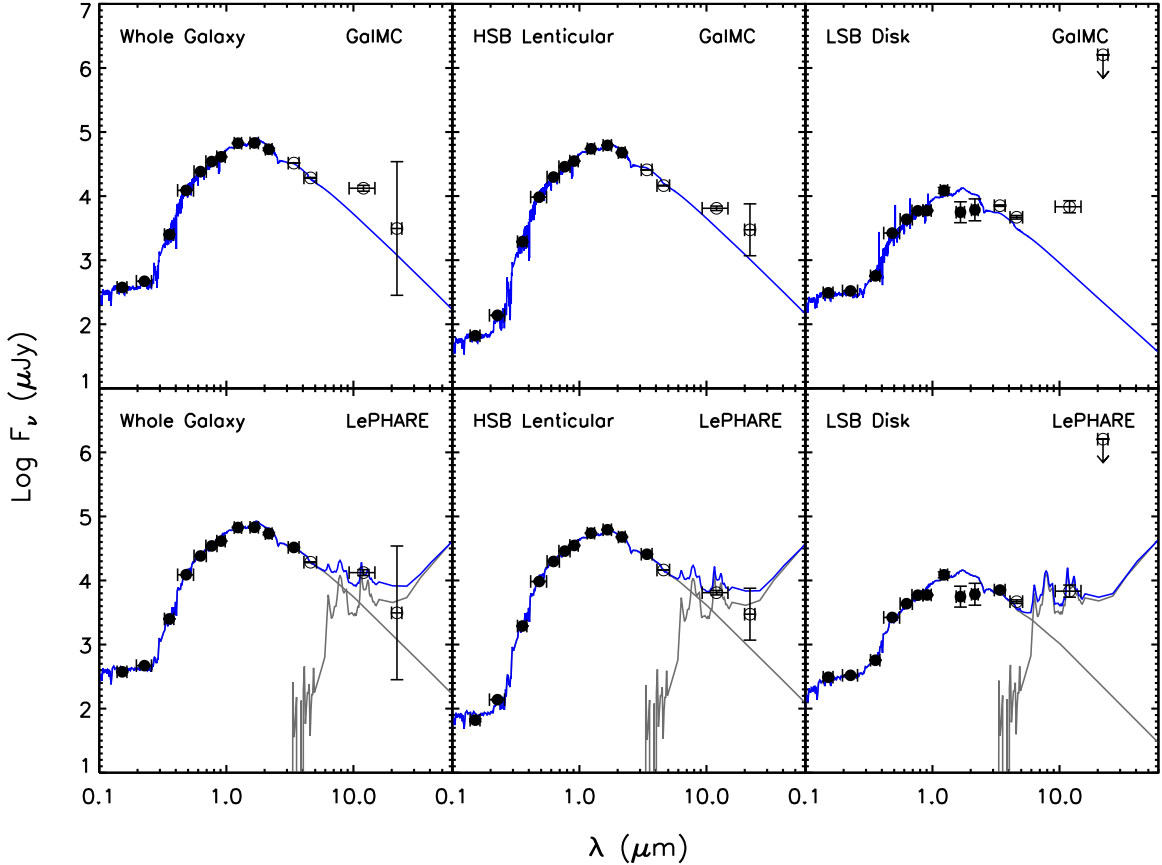}
	\caption{Photometric data from Table~\ref{tab-observations} overlaid with SED fits, shown in blue, for the whole galaxy, the HSB lenticular component, and the LSB disk. Data not used for fitting are marked with open circles.  \textit{Top row}: GalMC results.
	\textit{Bottom row}: LePHARE results, divided into the stellar and dusty components (grey lines).}
	\vspace{3mm}
	\label{fig-sed}
\end{figure*}


\subsection{GalMC}

GalMC utilizes a Markov-Chain Monte Carlo (MCMC) approach, avoiding the problems of Chi-square fitting routines, which can miss degeneracies or find local instead of global best fits.  GalMC fits over a range of 0.15 to 3~$\mu$m, and doesn't include low-energy dust physics.

We used Charlot and Bruzual 2007 stellar population synthesis models \citep{bruzual03} and adopted the \citet{salpeter55} initial mass function (IMF) with M$_\text{min} = 0.1$~\msun\ and M$_\text{max} = 100$~\msun.  We use the \citet{calzetti00} reddening law and account for absorption by the intergalactic medium using \citet{madau95}.  The metallicity was fixed at solar.  Five percent photometric errors were added in quadrature to the known errors in order to account for the error in absolute calibration.

We chose four free parameters: stellar mass, the time since the onset of star formation, E(B-V), and the exponential star formation timescale $\tau$, where $\text{SFR} \propto \exp(-t/\tau)$.  The current SFR is calculated as
\begin{equation}
\text{SFR} = \frac{e^{\text{A}/\tau}}{ e^{\text{A}/\tau} - 1}
\ \frac{\text{M}}{\tau}
\ e^{-\text{A}/\tau} ,
\end{equation}
where $A$ is the age (onset of star formation) and $M$ is the total mass.
We included nebular emission lines and nebular continuum only when fitting the spiral arms.  Although the HSB lenticular component has active star formation as well, the addition of emission lines to the HSB fits caused only a negligible change in the physical parameters. 

For the LSB disk component, $\tau$ was several Gyr, so we also considered a constant SFH. To determine whether the exponential or constant SFH model was best, we calculated their corrected Akaike information criterion \citep[AIC;][]{akaike74, hurvich89}.  The ratio of the exponential SFH AIC to the constant SFH AIC was $\sim$$10^{15}$, meaning that the exponential SFH is the superior model.  

A major concern when using MCMC methods is checking convergence.  We ran four chains from randomly chosen starting locations in parameter space to help ensure convergence \citep{acquaviva11}. We used CosmoMC's program GetDist \citep{lewis02} to analyze the chains. Since we have multiple chains, we use the \citeauthor{gelman92} $R$ statistic to test for convergence \citep{gelman92, brooks98}. All of our $R-1$ values were $\lesssim 0.02$, which shows convergence; the standard value for convergence is $R-1 < 0.1$.


\subsection{LePHARE}

The LePHARE FUV to near-IR SED fitting was done using a grid of \citet{bruzual03} stellar evolution models with a \citet{chabrier03} IMF with M$_\text{min} = 0.15$~\msun\ and M$_\text{max} = 120$~\msun. The grid was constructed from thirteen exponential SFH models with time scales ranging from 0.1 to 30~Gyr and metallicities of $Z_\odot$ and $0.4Z_\odot$. For each of these models, SEDs were computed for ages (time since formation) ranging from 0 to 13~Gyr. Dust attenuation is applied for 3 different extinction laws: SMC \citep{prevot84}, starburst galaxy \citep{calzetti00} and a power law with slope 0.9. We used thirteen discrete values of E(B-V) ranging from 0.0 to 0.6.
The code returns best fit values for physical parameters including stellar mass, SFR, specific SFR (SSFR), and age, as well as median values based on their probability distributions. The extinction law and E(B-V) are best fit values only.

\subsection{SED Modeling Results}

The model SEDs are overlaid on the photometric data in Figure~\ref{fig-sed}, and the associated physical parameters (stellar mass, age, reddening, SFR time scale, SFR, and SSFR) are given in Table~\ref{tab-sed}.  The processed mass (from GalMC modeling only) includes the current stellar mass plus the mass that has been processed through previous generations of stars.  The interpretation of age is ambiguous; mathematically, it's the time of the onset of star formation, but its physical meaning is difficult to determine, primarily due to degeneracies with other parameters.  The errors for the parameters account for statistical errors in the fitting process, and don't include systematic effects due to the models themselves.

\input{table4.tex}



Comparing the model parameters for each of the physical components, we find that the masses, SFRs, and time scales $\tau$ are logically consistent, i.e., combining the HSB and LSB model values yields something equivalent to the values found for the whole galaxy.
The parameters generated by GalMC and LePHARE are reasonably consistent, and we attribute most of the differences to the use of different stellar libraries.


We find that the stellar mass of the galaxy is approximately $8 \times 10^{10}$~\msun.
Both the GalMC and LePHARE values are consistent with the previous estimate of 
\mbox{$5.0^{+14.9}_{-3.8} \times 10^{10}~\msun$}
\citep{west10}, which used the \citet{bell03} stellar fitting models.  
It is important to note that the choice of IMF affects the stellar mass measurement.  Converting the LePHARE stellar mass to a \citet{salpeter55} IMF increases the mass by about 0.2~dex \citep{chabrier03}, making it even more similar to that measured by GalMC.
The models agree that the lenticular portion of the galaxy contributes about 80\% of the total stellar mass, and the extended LSB spiral arms provide the remaining 20\%.
The current stellar mass comprises 70\% of the total mass processed over the galaxy's lifetime.


The reddening within \ugc\ is low in both the HSB lenticular component and LSB spiral arms.  GalMC finds that all are consistent with zero.  LePHARE only measures modest reddening in the LSB arms, and none in the lenticular component.  Given the ongoing star formation within both the inner disk of the lenticular component and in the spiral arms, we would expect to see at least some reddening in both components.  However, since the HSB lenticular bulge and disk are modeled as a single component, the low-dust older stellar populations within the bulge likely dominate the reddening estimate.


The SFR of \ugc\ is extremely low in the HSB lenticular portion.  LePHARE computes the LSB disk SFR to be greater than that of the whole galaxy; this may be an artifact of combining two distinct populations into one model.  Taking the SFRs from GalMC, the LSB spiral arms dominate with rates of 0.2-0.3~\msun/yr, which is 85\% of the total SFR. 

The SFR can also be determined directly from the galaxy's UV flux.  We used the HI-based FUV attenuation correction from \citet{bigiel10}, which assumes (1)~MW-like attenuation, (2)~the FUV originates from the midplane so that only half of the dust/gas contributes to the attenuation, and (3)~$A_{FUV} /E(B-V) = 8.24$ \citep{wyder07}. 
Combined with the SFR law from \citet{salim07}, the LSB spiral arms have a SFR of $0.37^{+0.30}_{-0.17}$ \msun/yr, which is consistent with the SFRs found with both GalMC and LePHARE. In addition, we can find the SFR of the tight inner arms within the lenticular disk seen in Figure~\ref{fig-disksub}, since they are minimally contaminated by UV bulge flux.  Including only the flux between radii of 30 and 66 arcseconds (11 and 25~kpc; labeled in Figure~\ref{fig-profile}),
its SFR is $0.05^{+0.04}_{-0.02}$ \msun/yr, which is also very similar to the GalMC and LePHARE SFRs. 
The total UV-derived SFR of \ugc\ is therefore $0.42^{+0.30}_{-0.17}$ \msun/yr.


The exponential SFH time scale ($\tau$) is strongly degenerate with age, so one must be careful to not over-interpret either of them.  Therefore, we only make broad comparisons between components.  From both GalMC and LePHARE, we can deduce that the lenticular component formed rapidly compared to the LSB spiral arms, as both models suggest a shorter $\tau$.  Interestingly, both methods also suggest that the lenticular component is about 4~Gyr younger than the LSB spiral arms, so the spiral arms have likely been present in \ugc\ for a significant amount of time, which is difficult to explain by standard inside-out secular evolution.




\section{Evolution Through the Green Valley} \label{sec-green}

\begin{figure}
	\centering

	\includegraphics[trim = 35mm 130mm 75mm 32mm, clip=true, width=0.95\columnwidth]{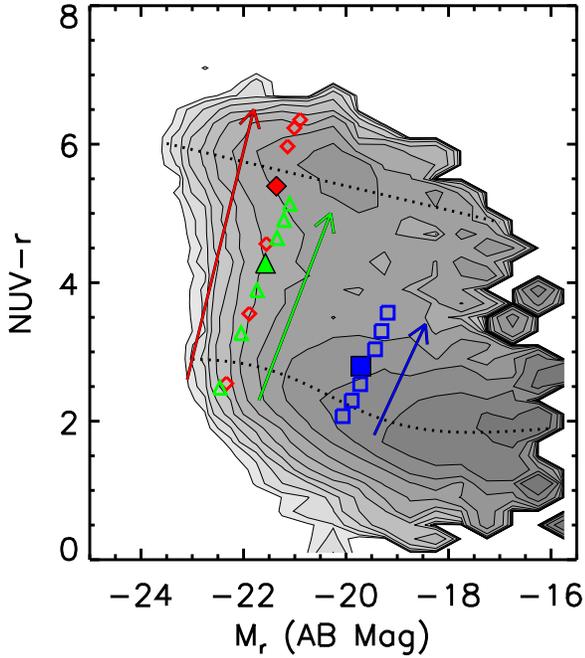}
	
	\caption{Galaxy CMD from \citet{wyder07} with the position of UGC~1382 (filled green triangle), which is in the green valley.  The positions of just the lenticular component (filled red diamond) and just the LSB spiral arms (filled blue square) are also marked.  The unfilled symbols refer to the location of each component at 1, 2, and 3~Gyr in the past and future, going from past to future as indicated by the arrows, assuming the exponentially declining SFH derived from our modeling. The dotted lines represent the locations of the red and blue sequences found in \citet{wyder07}.}
	\label{fig-cmd}
\end{figure}

The UV/optical color-magnitude diagram (CMD) of galaxies from \citet{wyder07} is an excellent diagnostic for separating red (passive) and blue (star-forming) galaxies due to both the long wavelength lever arm and utilizing UV as a direct tracer of SF.  The position of UGC~1382 is shown in Figure~\ref{fig-cmd}.  The lenticular component, considered alone, aligns with the red sequence of galaxies.  Once the LSB spiral arms are included, it shifts to the green valley.  This places UGC~1382 among the galaxies considered to be transitioning between the blue and red sequences.

The long exponential star formation timescale of $\tau \sim 2.5$ Gyr found by the SED fitting indicates that the spiral arms have been forming stars for a long time.  If the arms are old ($\sim$10~Gyr according to the SED modeling), it is plausible that the spiral arms have been present for most of the galaxy's history, and the galaxy is very slowly moving away from the blue sequence.

This is confirmed by simulating the galaxy's evolution.  
Individually for the lenticular component and LSB spiral arms, we set ages of 1, 2, and 3 Gyr into the past and future.  Using the associated GalMC exponential star formation time scale in Table~\ref{tab-sed}, we find the SFR and stellar mass at each age, with the assumption that the dust extinction doesn't evolve.
We extract the FUV, NUV, and $r$ photometry from the corresponding SEDs.  At each age, we add the extracted flux from the bulge and spiral arms to represent the entire galaxy.
The photometry for each of these is plotted in Figure~\ref{fig-cmd}, with the younger versions to the bottom-left and the aged versions to the upper-right.  
The evolution of the whole galaxy indicates that \ugc\ was much bluer and brighter in the past, so much so that it may have been in the blue sequence as recently as 2~Gyr ago.

The HSB lenticular component of the UGC~1382 is what one would expect -- old, nearly quenched, and relatively quiescent.  It resides on the red sequence (Figure~\ref{fig-cmd}), lies well below the galaxy main sequence (Figure~\ref{fig-ms}), and has a SSFR typical of non-star forming bulge systems.  The outer LSB spiral arm region would be a blue sequence galaxy by itself (Figure~\ref{fig-cmd}) and lies within $2\sigma$ of the galaxy main sequence (Figure~\ref{fig-ms}) with a SSFR for normal disks.

\begin{figure}
	\centering
	\includegraphics[trim = 125mm 50mm 15mm 15mm, clip=true, width=0.95\columnwidth, angle=180]{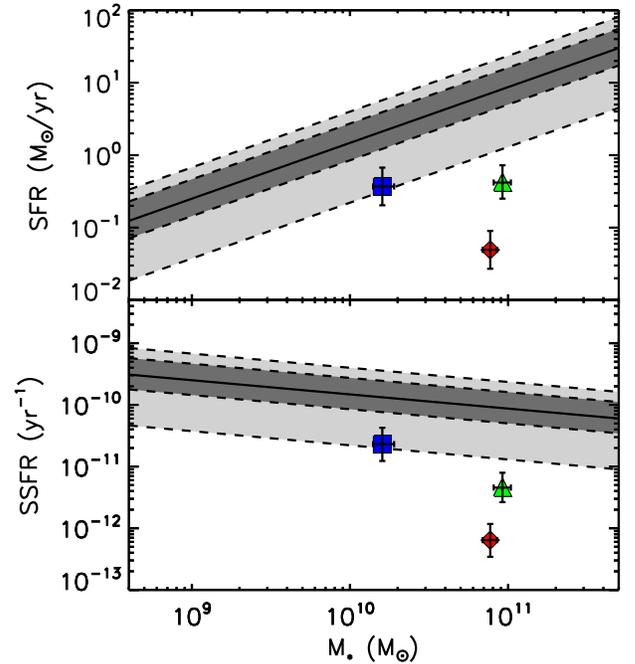}
	\caption{Position of UGC~1382 (green triangle) relative to the star-forming galaxy main sequence, displayed in terms of both SFR (top panel) and SSFR (bottom panel).  The grey regions are the $1\sigma$ and $2\sigma$ spread in the main sequence as found in \citet{elbaz07} for local SDSS galaxies ($0.015 \le z \le 0.1$).  The HSB lenticular (red diamond) and LSB spiral arms (blue square) of UGC~1382 are also included.  The UGC~1382 SFR is calculated from the FUV flux, though the Elbaz et al.\ relation is derived from H$\alpha$.}
	\vspace{2mm}
	\label{fig-ms}
\end{figure}

The fact that UGC~1382 is in the green valley implies that UGC~1382 is either evolving from the blue to red sequence as star formation shuts off (i.e., secular evolution) or that it was a red sequence galaxy that has been recently rejuvenated either by merger or accretion.  We believe that the former is more likely than the latter (see \SS\ref{sec-formation})

There are other early type systems with extended star formation as well. \citet{moffett12} describe a sample of low mass E/S0 galaxies that exhibit signatures of XUV disks and suggest such systems may be ubiquitous.  They hypothesize that these systems have recently acquired gas through mergers or cold accretion (i.e., rejuvenation). UGC~1382, if typical, does not support the idea of rejuvenation very well, although rejuvenation must certainly occur \citep[e.g., NGC 404,][]{thilker10}.  \citet{moffett12} also point out that this may be an important mechanism for disk growth in early type galaxies. Indeed, in the case of UGC~1382, the extended disk contains $\sim$20\% of the stellar mass.

The idea that UGC~1382 is a recently rejuvenated red sequence galaxy is unlikely to be true for three reasons: (1)~The outer LSB spiral arms, which are actively forming stars at a modest rate, appear to be at least as old as the central lenticular portion.  (2)~The outer spiral arms appear to be a continuation of the weak inner arms and disk and hence not a recently accreted system. (3)~The  HI is distributed in a large, uniformly rotating disk  and shows no obvious signs of significant interaction.  All of this points to the LSB spiral arms and disk not being a recent addition.

If, on the other hand, UGC~1382 is transitioning from the blue cloud to the red sequence, it is unlikely to complete the transformation very quickly.  At best, if the SFR is exponentially decaying, it may reach the red sequence within 3~Gyr. However, it is just as likely that, if undisturbed, the current modest SFR will continue at a constant rate.  Using our estimate of the total HI mass, if all of the HI gas were converted to stars at the GalMC and UV-derived SFRs, the gas depletion timescales are 
$64 ^{+181}_{-44}$~Gyr 
and 
$41 ^{+29}_{-17}$~Gyr, 
respectively.  Under a constant star formation scenario, UGC~1382 will effectively be a permanent  green valley resident.  The truth probably lies somewhere between.

\section{Formation Scenarios} \label{sec-formation}

The formation of GLSB galaxies is not well understood, and there are a number of possible formation scenarios. The most widely accepted theories involved either major collision perturbations  \citep[e.g.,][]{mapelli08}, the evolution of disk galaxies within a massive dark matter halo in isolated environments \citep[e.g.,][]{hoffman92}, or the tidal disruption and accretion of gas rich dwarf galaxies \citep[e.g.,][]{penarrubia06}.  The physical properties of UGC~1382 are more consistent with the dwarf tidal disruption scenario.

Most major galaxy interactions will likely result in the destruction of disks \citep{wilman13}. However, \cite{mapelli08} propose that collisional ring galaxies may be the precursors of GLSB galaxies. Their N-body simulations show that the expanding rings can redistribute both mass (stellar and gas) and angular momentum out to 100~kpc or more from the center of a galaxy while leaving a normal stellar bulge component at the center of (or slightly offset from) the system. The ring structure dominates for the first 100 to 200~Myr after the collision but will fade and become indistinguishable from the rest of the disk after 0.5 to 1~Gyr. As the disk expands, its surface density decreases by an order of magnitude. The surface brightness profiles of Malin~1, Malin~2, UGC~6614 and and NGC~7589 can be well matched to this model.

The collisional ring scenario is not well supported by UGC~1382 for two reasons.  First, the simulations predict relatively smooth LSB disks without the spiral arm structures that are obvious in UGC~1382 (as well as several other GLSB galaxies).  \citet{mapelli08} suggest that the lack of spiral arm structure in the models may be a consequence of numerical resolution.  
Second, because the ring structure would have an intense burst of star formation, one would expect the resultant extended LSB disk to be characterized by a stellar population with an average age younger than the bulge component. The UGC~1382 LSB disk is at best the same age as the bulge and very possibly older than the bulge. Furthermore, one would expect the exponential star formation timescale ($\tau$) for an expanding ring to be short, as the ring phase is less than $\sim$0.2~Gyr, but for the LSB disk of UGC~1382, $\tau \gtrsim3$~Gyr.

In contrast to a major collision,
another formation scenario proposes secular evolution in a low-density environment.
LSB disk galaxies of any size tend to be found in low-density environments near the outer parts of cosmological filaments and even in voids \citep{rosenbaum04, rosenbaum09}. Isolation from interactions and major mergers  may be a crucial element in the survival of LSB systems \citep{galaz11}.  GLSB galaxies have the additional property of being dominated by massive ($>10^{12} $~\msun) dark matter halos, which may protect against disk instabilities \citep[e.g.,][]{ostriker73, mayer04, das13}.
\citet{hoffman92} theorize that when rare large 3$\sigma$ density perturbations occur within voids, the cooling time is shorter than the dynamical time only in the central regions.  This leads to fast bulge formation while the outer regions form a thin, self-gravitating centrifugally-supported disk over a Hubble time. They conclude that this is a natural formation scenario for GLSB galaxies.  However, in this pure secular evolution theory, one would expect the LSB disk to have a mean age that is younger than the central HSB component. UGC~1382 conflicts with this scenario because the LSB disk is likely older than the HSB central region.

\citet{noguchi01} proposes another method to create a GLSB galaxy through secular evolution.  In this scenario, a normal HSB galaxy creates a bar, which induces radial mixing and moves material to the galaxy outskirts.  Over the course of several Gyr, the central surface brightness decreases by 1.5~mag/arcsec$^2$ and the disk scale length doubles.  This model predicts spiral arm features in the outer disk, which we observe, but it also requires the presence of a large bar, for which we see no evidence.  We therefore rule out this model as the way UGC~1382 formed.
As a final pathway for secular evolution, one can imagine a scenario in which the isolated halo initially formed a normal disk, then something triggers a disk instability, driving significant amounts of gas towards the center region.  This would form a HSB lenticular component in a relatively short period of time, while the outer portions continue their slower evolution.  However, what would limit the size of the lenticular disk to what we see today is mystery.

The final option we consider for the formation of a GLSB extended disk is the tidal disruption and accretion of dwarf galaxies. In an effort to explain the extended disk of M31, \citet{penarrubia06} model the accretion of co-planar dwarf galaxies and find that the mechanism can generate low surface brightness exponential disks as large as $r=200$~kpc depending on the stellar concentration of the accreted system. Although it depends on the initial orbit and stellar density of the satellite, an exponential disk about the host can develop as quickly as 3~Gyr after accretion for a 2:1 mass ratio.  Furthermore, spiral-like features in the extended disk are also predicted.  Another prediction of the model is that the extended LSB disk will have a circular velocity 30-50~km/s lower than the host if the initial satellite orbit was circular and 100-200~km/s lower if initially eccentric. The rotation curve of UGC~1382 hints at a modest turnover in the outer regions of $44 \pm16$~km/s relative to the peak value. Higher precision kinematic information, especially of the stellar component, is required to confirm this difference.

If the LSB disk of UGC~1382 originates from tidal disruption of dwarf galaxies, the model also needs to explain the high HI gas mass in the outer regions. Only 10\% of the neutral gas of UGC~1382 resides inside the central HSB lenticular component. This corresponds to a gas mass fraction $f_g = M_\text{gas}/(M_\text{gas} + M_*) = 0.03$. For the LSB disk, $f_g = 0.58$. Although the LSB disk is gas-rich, it falls within the typical range of LSB dwarf galaxies, which are the most gas rich systems known with $f_g = 0.4$ to 0.8 \citep{schombert01}. If the accretion event was 3~Gyr ago, following the star formation history of the LSB disk modeled in Section~\ref{sec-model}, the accreted material would have had $f_g = 0.63$.  Hence the accretion of a single large-mass or several moderate-mass systems are sufficient to provide the stellar and gas material found in the extended disk of UGC~1382. 
LSB dwarfs are preferentially located in low density environment \citep{rosenbaum04, rosenbaum09} such as that of UGC~1382, so accretion of one or more would be feasible. By allowing independent evolutionary histories for the HSB central component and the extended LSB disk, the relative ages of the stellar populations can be easily explained. That is, the age of LSB disk should reflect the age of the accreted satellite(s), which can be older than the central HSB disk.
It is interesting to consider that one of the projected systems, SDSS~J015436.03-000922.7, which appears to be a low mass bulge-type system, may be a candidate for a remnant core of a LSB dwarf galaxy.  If the HI disk truly has a slight warp, it may be attributed to the accretion of a dwarf system.
The only discrepancy with the dwarf accretion scenario is in the morphology of UGC~1382.  We have presented imaging evidence that the inner HSB lenticular spiral arms may connect directly to the outer LSB spiral arms (see Figure~\ref{fig-disksub}); this suggests that the LSB and HSB components might not be kinematically distinct.  If so, this may contradict an accretion scenario.

Because there are no strong inconsistencies between UGC~1382 and the dwarf satellite accretion build-up of its LSB disk, we favor this formation scenario.  Further observations exploring the stellar and gas metallicity and kinematics of the HSB component and the LSB disk in detail will help confirm or refute this formation mechanism.


\section{Summary and Conclusions} \label{sec-summary}

We have presented observations and analysis of UGC~1382, a giant low surface brightness (GLSB) galaxy that has previously been mistaken for an elliptical galaxy.  Below are our conclusions about this unique system.

(i) Morphologically, UGC~1382 can be described as being composed of a HSB lenticular galaxy surrounded by a LSB disk.  This LSB disk has a spiral arm structure that appears to continue into the lenticular component.
The absolute size of UGC~1382 is extraordinary: optical and UV light is detected to a radius of 80~kpc, and the HI disk extends to a radius of 110~kpc.  This places UGC~1382 among the largest disk galaxies currently known.

(ii) There are many lines of evidence that UGC~1382 is a GLSB galaxy.
The disk has an effective radius of $r_e = 38$~kpc, an exponential scale length of 28.5~kpc, and a central surface brightness of 26.2~mag/arcsec$^2$, which is comparable to those of Malin~1 (the prototypical GLSB galaxy) and Malin~2.  It contains $1.7 \times 10^{10}$~\msun\ of HI gas at a very low surface density ($< 3$~\msun/pc$^2$).  The SFR efficiency is everywhere quite low and most similar to the far outer regions of normal spiral galaxies.

(iii) UGC~1382 resides in a low-density environment; it has a dwarf system/remnant embedded in its LSB disk, and the nearest large galaxy is about 1.6~Mpc away.
Like other GLSB galaxies, UGC~1382 is dark-matter dominated with a dark matter fraction of 0.95, and it has a dark matter halo mass of $2\times 10^{12}$~\msun.  It is therefore unsurprising that we find no evidence for recent major galaxy interactions.  It is likely that the LSB disk component of UGC~1382 could only exist in this type of environment.

(iv) From modeling the SEDs of both the inner lenticular component and outer LSB spiral arms, we have determined that the total stellar mass of UGC~1382 is $8 \times 10^{10}$~\msun, 20\% of which is in the LSB component.  Both components are nearly dust-free.  The overall SFR is $\sim$0.4~\msun/yr, 85\% of which is in the LSB disk.  The LSB disk is about 4~Gyr older than the central lenticular component and has a characteristic star formation timescale that is significantly longer; this makes it unlikely that the disk is a recent acquisition. 
 
(v) UGC~1382 resides in the ``green valley" of the galaxy CMD.  Based on the SED modeling result that the star formation is exponentially declining, the galaxy appears to be transitioning from the blue to the red sequence.  However, with its low SFR and huge HI reservoir, it has the capacity to stay in the green valley for several Hubble times.

(vi) The properties of UGC~1382 are most consistent with the formation scenario following \citet{penarrubia06}.  In this scenario, the core lenticular galaxy accretes one (or more) gas-rich dwarf galaxies, which are responsible for both forming the disk and for providing the fuel for the subsequent and ongoing low-level star formation.  This quantitatively agrees with our observations of a gas-rich disk that is older and and has a longer star-formation time scale than the lenticular core, as well as with the presence of a tiny dwarf galaxy in the LSB disk.

The detection of the extended, LSB stellar component surrounding UGC1382 and subsequent classification of this system as a GLSB galaxy has implications on our understanding of other seemingly ``normal" early-type galaxies.  The increasing availability of sensitive optical, UV, and HI observations of early-type galaxies may reveal additional cases similar to UGC1382.  A census of the low-surface-brightness stellar and gas content of early-type galaxies would help place interesting constraints on star formation properties in bulge-dominated galaxies as well as the importance of different processes in the evolution of galaxies.


\ \\

\acknowledgements

We thank the referee for helpful comments that improved this paper.
MHS acknowledges support from NASA grant NNX12AE19G.
LMY acknowledges support from NSF AST-1109803 and thanks ASIAA for their hospitality during a sabbatical visit.
This publication makes use of data products from the Two Micron All Sky Survey, which is a joint project of the University of Massachusetts and the Infrared Processing and Analysis Center/California Institute of Technology, funded by the National Aeronautics and Space Administration and the National Science Foundation.
This research has made use of NASA's Astrophysics Data System.
This research has made use of the NASA/IPAC Extragalactic Database (NED) which is operated by the Jet Propulsion Laboratory, California Institute of Technology, under contract with the National Aeronautics and Space Administration.
We thank the Research Computer and Cyberinfrastructure Unit of Information Technology Services at The Pennsylvania State University for providing computational support and resources. In particular, we appreciate the very helpful William Brouwer.
The Institute for Gravitation and the Cosmos is supported by the Eberly College of Science and the Office of the Senior Vice President for Research at the Pennsylvania State University.
Observations are based in part on data obtained at the duPont 2.5m telescope at the Las Campanas Observatories of the Carnegie Institution for Science.

\bibliographystyle{apj}

\bibliography{apj-jour,bibliography_file}



\end{document}

%% file: table1.tex
\begin{deluxetable*}{l cc ccc}
\tablecaption{Summary of UV to Mid-IR Observations of UGC~1382 \label{tab-observations}}
\tablewidth{0pt}
\tabletypesize{\small}
\tablehead{
\colhead{} & \colhead{} & \colhead{}
& \multicolumn{3}{c}{Photometry (AB Mag)}
\\
\cline{4-6}
\colhead{} & \colhead{Wavelength} & \colhead{5$\sigma$ Limiting} 
& \colhead{Total} & \colhead{HSB Lenticular} & \colhead{LSB Spiral Arms}
\\
\colhead{Band} & \colhead{($\mu$m)} & \colhead{Magnitude (AB)} 
& \colhead{$r < 180''$} & \colhead{$r < 66''$} & \colhead{$66'' < r < 180''$}
\\
\colhead{(1)} & \colhead{(2)} & \colhead{(3)} & \colhead{(4)} & \colhead{(5)} & \colhead{(6)}
}
\startdata

\galex\ FUV & 0.1516 & 22.77 & 17.68 $\pm$ 0.02 & 19.57 $\pm$ 0.05 & 17.89 $\pm$ 0.03  \\
\galex\ NUV & 0.2267 & 23.01 & 17.44 $\pm$ 0.02 & 18.77 $\pm$ 0.02 & 17.82 $\pm$ 0.03  \\
SDSS $u$ & 0.3557 & 23.64 & 15.54 $\pm$ 0.01 & 15.82 $\pm$ 0.01 & 17.15 $\pm$ 0.01  \\
SDSS $g$ & 0.4825 & 24.58 & 13.78 $\pm$ 0.01 & 14.04 $\pm$ 0.01 & 15.45 $\pm$ 0.01  \\
SDSS $r$ & 0.6261 & 24.01 & 13.02 $\pm$ 0.01 & 13.23 $\pm$ 0.01 & 14.88 $\pm$ 0.01  \\
SDSS $i$ & 0.7672 & 23.53 & 12.61 $\pm$ 0.01 & 12.81 $\pm$ 0.01 & 14.53 $\pm$ 0.01  \\
SDSS $z$ & 0.9097 & 22.16 & 12.41 $\pm$ 0.01 & 12.57 $\pm$ 0.01 & 14.51 $\pm$ 0.01  \\
2MASS J & 1.235 & 18.18 & 11.86 $\pm$ 0.02 & 12.08 $\pm$ 0.01 & 13.70 $\pm$ 0.12  \\
2MASS H & 1.662 & 17.18 & 11.84 $\pm$ 0.03 & 11.94 $\pm$ 0.01 & 14.55 $\pm$ 0.41  \\
2MASS K$_s$ & 2.159 & 16.57 & 12.08 $\pm$ 0.05 & 12.21 $\pm$ 0.02 & 14.45 $\pm$ 0.43  \\
WISE w1 & 3.4 & 19.49 & 12.62 $\pm$ 0.01 & 12.88 $\pm$ 0.01 & 14.28 $\pm$ 0.03  \\
WISE w2 & 4.6 & 19.45 & 13.19 $\pm$ 0.02 & 13.50 $\pm$ 0.01 & 14.73 $\pm$ 0.07  \\
WISE w3 & 12 & 18.03 & 13.59 $\pm$ 0.11 & 14.38 $\pm$ 0.09 & 14.32 $\pm$ 0.24  \\
WISE w4 & 22 & 16.21 & 15.16 $\pm$ 2.61 & 15.22 $\pm$ 1.01 & $ >8.39 $  \\

\enddata
\tablecomments{These data are not corrected for foreground galactic extinction.}
\end{deluxetable*}

%% file: table2.tex
\begin{deluxetable*}{l c}
\tablecaption{Selected Properties of UGC~1382 \label{tab-summary}}
\tablewidth{0pt}
\tabletypesize{\footnotesize}
\tablehead{
\colhead{Quantity} & \colhead{Value}
}
\startdata

\hline
RA (J2000) & 28.671011$^\circ$ \\
Dec (J2000) & -0.143342$^\circ$ \\
Distance & 80 Mpc \\
UV/Optical Radius & 80 kpc \\
HI Radius & 110 kpc \\
SFR & $0.42 ^{+0.30} _{-0.17}$ \msun/yr  \\
Radial Velocity & $5591 \pm 2$ km/s \\
$V_\text{rot}$ at 110 kpc & 280 km/s \\
HI Mass & $1.7 (\pm 0.1) \times 10^{10}$ \msun \\
Stellar Mass ($r < 70$~kpc) & $8 \times 10^{10}$ \msun \\
Dynamical Mass ($r < 110$~kpc) & $2 \times 10^{12}$ \msun \\
Dark Matter Fraction ($r < 110$~kpc) & 0.95 \\
r-band Bulge-to-Disk Ratio (Lenticular Component) & 0.70 \\

\enddata


\end{deluxetable*}

%% file: table3.tex
\begin{deluxetable*}{l ccc}
\tablecaption{Summary of \ser\ Fits for UGC~1382 \label{tab-sersic}}
\tablewidth{0pt}
\tabletypesize{\footnotesize}
\tablehead{
\colhead{} & \multicolumn{3}{c}{Morphological Component} \\
\cline{2-4}
\colhead{Quantity} & \colhead{Bulge} & \colhead{Inner Disk} & \colhead{Outer LSB Arms}
}
\startdata

\ser\ Index, $n$ & 3.5 & 1.4 & 0.5 \\
Effective Radius (kpc), $r_e$ & 1.3 ($3.4''$) & 6.0 ($16''$) & 38 ($100''$) \\
Central $r$-band Surface & \multirow{2}{*}{17.5} & \multirow{2}{*}{20.1} & \multirow{2}{*}{25.8} \\
\ \ \ Brightness (mag/arcsec$^2$) & & & \\
Total $m_r$ (mag) & 14.3 & 13.9 & 14.6 \\

\enddata


\end{deluxetable*}

%% file: table4.tex
\begin{deluxetable*}{l ccccccc}
\tablecaption{Physical Parameters from SED Fitting \label{tab-sed}}
\tablewidth{0pt}
\tabletypesize{\footnotesize}
\tablehead{
\colhead{Morphological} & \colhead{Processed Mass} & \colhead{Stellar Mass} 
& \colhead{Age} & \colhead{E(B-V)} & \colhead{$\tau$} & \colhead{SFR} 
& \colhead{Log sSFR}
\\
\colhead{Component} & \colhead{($10^{10}$ \msun)} & \colhead{($10^{10}$ \msun)}
& \colhead{(Gyr)} & \colhead{(mag)} & \colhead{(Gyr)} & \colhead{(\msun/yr)} 
& \colhead{(yr$^{-1}$)}
}
\startdata

\multicolumn{2}{l}{\textit{GalMC}} & & & & & & \\ 

Whole Galaxy & 13.0 $ ^{ +1.7} _{ -1.6} $ & 9.2 $ ^{ +1.2} _{ -1.1} $ & 6.3 $ ^{ +1.1} _{ -1.0} $ & 0.006 $ ^{ +0.001}_{ -0.006} $ & 1.07 $ ^{ +0.22}_{ -0.16} $ & 0.26$ ^{ + 0.59}_{ - 0.19} $ & -11.44$ ^{ + 0.51}_{ - 0.41} $ \\
HSB Lenticular  & 10.9 $ ^{ +1.3} _{ -1.1} $ & 7.7 $ ^{ +0.8} _{ -0.8} $ & 5.5 $ ^{ +0.8} _{ -0.7} $ & 0.007 $ ^{ +0.002}_{ -0.007} $ & 0.70 $ ^{ +0.11}_{ -0.08} $ & 0.04$ ^{ + 0.12}_{ - 0.03} $ & -12.13$ ^{ + 0.60}_{ - 0.44} $ \\
LSB Spiral Arms & 2.3 $ ^{ +0.4} _{ -0.3} $ & 1.6 $ ^{ +0.3} _{ -0.2} $ & 9.3 $ ^{ +2.5} _{ -1.9} $ & 0.010 $ ^{ +0.002}_{ -0.010} $ & 2.66 $ ^{ +0.93}_{ -0.56} $ & 0.22$ ^{ + 0.39}_{ - 0.15} $ & -10.79$ ^{ + 0.44}_{ - 0.40} $ \\

\hline
\multicolumn{2}{l}{\textit{LePHARE}} & & & & & & \\ 

Whole Galaxy & \nodata  & 6.7 $ ^{ +3.4} _{ -0.8} $ & 12.5 $ ^{ +1.1} _{ -1.0} $ & 0.000 & 2.64 $ ^{ +0.59}_{ -0.38} $ & 0.18$ ^{ + 0.08}_{ - 0.04} $ & -11.62$ ^{ + 0.09}_{ - 0.05} $ \\
HSB Lenticular  & \nodata  & 4.3 $ ^{ +2.2} _{ -0.5} $ & 7.9 $ ^{ +0.6} _{ -0.6} $ & 0.000 & 1.14 $ ^{ +0.17}_{ -0.13} $ & 0.03$ ^{ + 0.01}_{ - 0.01} $ & -12.19$ ^{ + 0.06}_{ - 0.07} $ \\
LSB Spiral Arms & \nodata  & 1.1 $ ^{ +0.3} _{ -0.4} $ & 11.3 $ ^{ +1.7} _{ -1.8} $ & 0.100 & 5.52 $ ^{ +4.57}_{ -2.01} $ & 0.30$ ^{ + 0.20}_{ - 0.11} $ & -10.51$ ^{ + 0.12}_{ - 0.20} $ \\

\enddata

\tablecomments{The physical parameters found for \ugc\ and its components using the GalMC and LePHARE SED fitting routines.  1$\sigma$ errors are included when calculated by each routine.}

\end{deluxetable*}